
\overfullrule=0pt


\chardef\tempcat=\the\catcode`\@ \catcode`\@=11
\def\cyracc{\def\u##1{\if \i##1\accent"24 i%
    \else \accent"24 ##1\fi }}
\newfam\cyrfam
\font\tencyr=wncyr10
\newfam\cyrfam

\def\cyr{\fam\cyrfam\tencyr\cyracc}


\input epsf

\def\figin{\epsfcheck\figin}\def\figins{\epsfcheck\figins}
\def\epsfcheck{\ifx\epsfbox\UnDeFiNeD
\message{(NO epsf.tex, FIGURES WILL BE IGNORED)}
\gdef\figin##1{\vskip2in}\gdef\figins##1{\hskip.5in}
\else\message{(FIGURES WILL BE INCLUDED)}%
\gdef\figin##1{##1}\gdef\figins##1{##1}\fi}
\def\DefWarn#1{}
\def\figinsert{\goodbreak\topinsert}
\def\ifig#1#2#3#4{\DefWarn#1\xdef#1{fig.~\the\figno}
\writedef{#1\leftbracket fig.\noexpand~\the\figno}%
\figinsert\figin{\centerline{\epsfxsize=#3mm \epsfbox{#2}}}
\bigskip\medskip\centerline{\vbox{\baselineskip12pt
\advance\hsize by -1truein\noindent\footnotefont{\sl Fig.~\the\figno:}\sl\ #4}}
\bigskip\endinsert\noindent\global\advance\figno by1}

\def\Figx#1#2#3{
\bigskip
\vbox{\centerline{\epsfxsize=#1 cm \epsfbox{NOpic#2.eps}}
\centerline{{\bf Fig.\the\figno} #3}}\bigskip\global\advance\figno by1}

\def\Figy#1#2#3{
\bigskip
\vbox{\centerline{\epsfysize=#1 cm \epsfbox{NOpic#2.eps}}
\centerline{{\bf Fig.\the\figno} #3}}\bigskip\global\advance\figno by1}
\newcount\figno
 \figno=1
 \def\fig#1#2#3{
 \par\begingroup\parindent=0pt\leftskip=1cm\rightskip=1cm\parindent=0pt
 \baselineskip=11pt
 \global\advance\figno by 1
 \midinsert
 \epsfxsize=#3
 \centerline{\epsfbox{#2}}
 \vskip 12pt
 {\bf Fig.\ \the\figno: } #1\par
 \endinsert\endgroup\par
 }
 \def\figlabel#1{\xdef#1{\the\figno}}
 \def\encadremath#1{\vbox{\hrule\hbox{\vrule\kern8pt\vbox{\kern8pt
 \hbox{$\displaystyle #1$}\kern8pt}
 \kern8pt\vrule}\hrule}}


\def\unlockat{\catcode`\@=11}

\unlockat

\global\newcount\secno \global\secno=0
\global\newcount\prono \global\prono=0
\def\newsec#1{\vfill\eject\global\advance\secno by1\message{(\the\secno. #1)}
\global\subsecno=0\global\subsubsecno=0
\global\deno=0\global\teno=0
\eqnres@t\noindent
{\titlefont\the\secno. #1}
\writetoca{{\bf\secsym} {\rm #1}}\par\nobreak\medskip\nobreak}
\global\newcount\subsecno \global\subsecno=0
\def\subsec#1{\global\advance\subsecno
by1\message{(\secsym\the\subsecno. #1)}
\ifnum\lastpenalty>9000\else\bigbreak\fi
\global\subsubsecno=0
\global\deno=0
\global\teno=0
\noindent{\bf\secsym\the\subsecno. #1}
\writetoca{\bf \string\quad {\secsym\the\subsecno.} {\it  #1}}
\par\nobreak\medskip\nobreak}
\global\newcount\subsubsecno \global\subsubsecno=0
\def\subsubsec#1{\global\advance\subsubsecno by1
\message{(\secsym\the\subsecno.\the\subsubsecno. #1)}
\ifnum\lastpenalty>9000\else\bigbreak\fi
\noindent\quad{\bf \secsym\the\subsecno.\the\subsubsecno.}{\ \sl \ #1}
\writetoca{\string\qquad\bf { \secsym\the\subsecno.\the\subsubsecno.}{\sl  \ #1}}
\par\nobreak\medskip\nobreak}

\global\newcount\deno \global\deno=0
\def\de#1{\global\advance\deno by1
\message{(\bf Definition\quad\secsym\the\subsecno.\the\deno #1)}
\ifnum\lastpenalty>9000\else\bigbreak\fi
\noindent{\bf Definition\quad\secsym\the\subsecno.\the\deno}{#1}
\writetoca{\string\qquad{\secsym\the\subsecno.\the\deno}{#1}}}

\global\newcount\prono \global\prono=0
\def\pro#1{\global\advance\prono by1
\message{(\bf Proposition\quad\secsym\the\subsecno.\the\prono 
)}
\ifnum\lastpenalty>9000\else\bigbreak\fi
\noindent{\bf Proposition\quad
\the\prono\quad}{\ninepoint #1}
}

\global\newcount\teno \global\prono=0
\def\te#1{\global\advance\teno by1
\message{(\bf Theorem\quad\secsym\the\subsecno.\the\teno #1)}
\ifnum\lastpenalty>9000\else\bigbreak\fi
\noindent{\bf Theorem\quad\secsym\the\subsecno.\the\teno}{#1}
\writetoca{\string\qquad{\secsym\the\subsecno.\the\teno}{#1}}}
\def\subsubseclab#1{\DefWarn#1\xdef #1{\noexpand\hyperref{}{subsubsection}%
{\secsym\the\subsecno.\the\subsubsecno}%
{\secsym\the\subsecno.\the\subsubsecno}}%
\writedef{#1\leftbracket#1}\wrlabeL{#1=#1}}

\def\unredoffs{} \def\redoffs{\voffset=-.40truein\hoffset=-.40truein}
\def\speclscape{}

\newbox\leftpage \newdimen\fullhsize \newdimen\hstitle \newdimen\hsbody
\tolerance=1000\hfuzz=2pt

\catcode`\@=11
\def\bigans{b }
\def\answ{b }

\ifx\answ\bigans\message{(This will come out unreduced.}
\magnification=1200\unredoffs\baselineskip=16pt plus 2pt minus 1pt
\hsbody=\hsize \hstitle=\hsize

\else\message{(This will be reduced.} \let\l@r=L
\magnification=1200\baselineskip=16pt plus 2pt minus 1pt \vsize=7truein
\redoffs \hstitle=8truein\hsbody=4.75truein\fullhsize=10truein\hsize=\hsbody
\output={\ifnum\pageno=0

   \shipout\vbox{{\hsize\fullhsize\makeheadline}
     \hbox to \fullhsize{\hfill\pagebody\hfill}}\advancepageno
   \else
   \almostshipout{\leftline{\vbox{\pagebody\makefootline}}}\advancepageno
   \fi}
\def\almostshipout#1{\if L\l@r \count1=1 \message{[\the\count0.\the\count1]}
       \global\setbox\leftpage=#1 \global\let\l@r=R
  \else \count1=2
   \shipout\vbox{\speclscape{\hsize\fullhsize\makeheadline}
       \hbox to\fullhsize{\box\leftpage\hfil#1}}  \global\let\l@r=L\fi}
\fi

\newcount\yearltd\yearltd=\year\advance\yearltd by -2000

\def\Title#1#2{\nopagenumbers
\abstractfont\hsize=\hstitle\rightline{#1}%
\vskip 5pt\centerline{\titlefont #2}\abstractfont\vskip .5in\pageno=0}
\def\Date#1{\vfill\centerline{#1}\tenpoint\supereject\global\hsize=\hsbody%
\footline={\hss\tenrm\folio\hss}}


\def\draftmode{\message{ DRAFTMODE }\def\draftdate{{\rm preliminary draft:
\number\month/\number\day/\number\yearltd\ \ \hourmin}}%

\writelabels\baselineskip=20pt plus 2pt minus 2pt
  {\count255=\time\divide\count255 by 60 \xdef\hourmin{\number\count255}
   \multiply\count255 by-60\advance\count255 by\time
   \xdef\hourmin{\hourmin:\ifnum\count255<10 0\fi\the\count255}}}

\def\nolabels{\def\wrlabeL##1{}\def\eqlabeL##1{}\def\reflabeL##1{}}
\def\writelabels{\def\wrlabeL##1{\leavevmode\vadjust{\rlap{\smash%
{\line{{\escapechar=` \hfill\rlap{\sevenrm\hskip.03in\string##1}}}}}}}%
\def\eqlabeL##1{{\escapechar-1\rlap{\sevenrm\hskip.05in\string##1}}}%
\def\reflabeL##1{\noexpand\llap{\noexpand\sevenrm\string\string\string##1}}}
\nolabels
%


\global\newcount\secno \global\secno=0
\global\newcount\meqno
\global\meqno=1
\def\eqnres@t{\xdef\secsym{\the\secno.}\global\meqno=1
\bigbreak\bigskip}
\def\sequentialequations{\def\eqnres@t{\bigbreak}}
\def\appendix#1#2{\vfill\eject\global\meqno=1\global\subsecno=0\xdef\secsym{\hbox{#1.}}
\bigbreak\bigskip\noindent{\bf Appendix #1. #2}\message{(#1. #2)}
\writetoca{Appendix {#1.} {#2}}\par\nobreak\medskip\nobreak}

\def\eqnn#1{\xdef #1{(\secsym\the\meqno)}\writedef{#1\leftbracket#1}%
\global\advance\meqno by1\wrlabeL#1}
\def\eqna#1{\xdef #1##1{\hbox{$(\secsym\the\meqno##1)$}}
\writedef{#1\numbersign1\leftbracket#1{\numbersign1}}%
\global\advance\meqno by1\wrlabeL{#1$\{\}$}}
\def\eqn#1#2{\xdef #1{(\secsym\the\meqno)}\writedef{#1\leftbracket#1}%
\global\advance\meqno by1$$#2\eqno#1\eqlabeL#1$$}

\newskip\footskip\footskip14pt plus 1pt minus 1pt

\def\footnotefont{\ninepoint}\def\f@t#1{\footnotefont #1\@foot}
\def\f@@t{\baselineskip\footskip\bgroup\footnotefont\aftergroup\@foot\let\next}
\setbox\strutbox=\hbox{\vrule height9.5pt depth4.5pt width0pt}
\global\newcount\ftno \global\ftno=0
\def\foot{\global\advance\ftno by1\footnote{$^{\the\ftno}$}}

\newwrite\ftfile
\def\footend{\def\foot{\global\advance\ftno by1\chardef\wfile=\ftfile
$^{\the\ftno}$\ifnum\ftno=1\immediate\openout\ftfile=foots.tmp\fi%
\immediate\write\ftfile{\noexpand\smallskip%
\noexpand\item{f\the\ftno:\ }\pctsign}\findarg}%
\def\footatend{\vfill\eject\immediate\closeout\ftfile{\parindent=20pt
\centerline{\bf Footnotes}\nobreak\bigskip\input foots.tmp }}}
\def\footatend{}

\global\newcount\refno \global\refno=1
\newwrite\rfile
\def\ref{[\the\refno]\nref}
\def\nref#1{\xdef#1{[\the\refno]}\writedef{#1\leftbracket#1}%
\ifnum\refno=1\immediate\openout\rfile=refs.tmp\fi \global\advance\refno
by1\chardef\wfile=\rfile\immediate \write\rfile{\noexpand\item{#1\
}\reflabeL{#1\hskip.31in}\pctsign}\findarg}

\def\findarg#1#{\begingroup\obeylines\newlinechar=`\^^M\pass@rg}
{\obeylines\gdef\pass@rg#1{\writ@line\relax #1^^M\hbox{}^^M}%
\gdef\writ@line#1^^M{\expandafter\toks0\expandafter{\striprel@x #1}%
\edef\next{\the\toks0}\ifx\next\em@rk\let\next=\endgroup\else\ifx\next\empty%
\else\immediate\write\wfile{\the\toks0}\fi\let\next=\writ@line\fi\next\relax}}
\def\striprel@x#1{} \def\em@rk{\hbox{}}
\def\lref{\begingroup\obeylines\lr@f}
\def\lr@f#1#2{\gdef#1{\ref#1{#2}}\endgroup\unskip}
\def\semi{;\hfil\break}
\def\addref#1{\immediate\write\rfile{\noexpand\item{}#1}}

\def\startrefs#1{\immediate\openout\rfile=refs.tmp\refno=#1}
\def\xref{\expandafter\xr@f}\def\xr@f[#1]{#1}
\def\refs#1{\count255=1[\r@fs #1{\hbox{}}]}
\def\r@fs#1{\ifx\und@fined#1\message{reflabel \string#1 is undefined.}%
\nref#1{need to supply reference \string#1.}\fi%
\vphantom{\hphantom{#1}}\edef\next{#1}\ifx\next\em@rk\def\next{}%
\else\ifx\next#1\ifodd\count255\relax\xref#1\count255=0\fi%
\else#1\count255=1\fi\let\next=\r@fs\fi\next}




%
\catcode`\@=12 
%
\edef\tfontsize{\ifx\answ\bigans scaled\magstep3\else scaled\magstep4\fi}
\font\titlerm=cmr10 \tfontsize \font\titlerms=cmr7 \tfontsize
\font\titlermss=cmr5 \tfontsize \font\titlei=cmmi10 \tfontsize
\font\titleis=cmmi7 \tfontsize \font\titleiss=cmmi5 \tfontsize
\font\titlesy=cmsy10 \tfontsize \font\titlesys=cmsy7 \tfontsize
\font\titlesyss=cmsy5 \tfontsize \font\titleit=cmti10 \tfontsize
\skewchar\titlei='177 \skewchar\titleis='177 \skewchar\titleiss='177
\skewchar\titlesy='60 \skewchar\titlesys='60 \skewchar\titlesyss='60
\def\titlefont{\def\rm{\fam0\titlerm}
\textfont0=\titlerm \scriptfont0=\titlerms \scriptscriptfont0=\titlermss
\textfont1=\titlei \scriptfont1=\titleis \scriptscriptfont1=\titleiss
\textfont2=\titlesy \scriptfont2=\titlesys \scriptscriptfont2=\titlesyss
\textfont\itfam=\titleit
\def\it{\fam\itfam\titleit}\rm}
 \ifx\answ\bigans\else scaled\magstep1\fi
\ifx\answ\bigans\def\abstractfont{\tenpoint}\else \font\abssl=cmsl10 scaled
\magstep1 \font\absrm=cmr10 scaled\magstep1 \font\absrms=cmr7
scaled\magstep1 \font\absrmss=cmr5 scaled\magstep1 \font\absi=cmmi10
scaled\magstep1 \font\absis=cmmi7 scaled\magstep1 \font\absiss=cmmi5
scaled\magstep1 \font\abssy=cmsy10 scaled\magstep1 \font\abssys=cmsy7
scaled\magstep1 \font\abssyss=cmsy5 scaled\magstep1 \font\absbf=cmbx10
scaled\magstep1 \skewchar\absi='177 \skewchar\absis='177
\skewchar\absiss='177 \skewchar\abssy='60 \skewchar\abssys='60
\skewchar\abssyss='60
\def\abstractfont{\def\rm{\fam0\absrm}
\textfont0=\absrm \scriptfont0=\absrms \scriptscriptfont0=\absrmss
\textfont1=\absi \scriptfont1=\absis \scriptscriptfont1=\absiss
\textfont2=\abssy \scriptfont2=\abssys \scriptscriptfont2=\abssyss
\textfont\itfam=\bigit \def\it{\fam\itfam\bigit}\def\footnotefont{\tenpoint}%
\textfont\slfam=\abssl \def\sl{\fam\slfam\abssl}%
\textfont\bffam=\absbf \def\bf{\fam\bffam\absbf}\rm}\fi
\def\tenpoint{\def\rm{\fam0\tenrm}
\textfont0=\tenrm \scriptfont0=\sevenrm \scriptscriptfont0=\fiverm
\textfont1=\teni  \scriptfont1=\seveni  \scriptscriptfont1=\fivei
\textfont2=\tensy \scriptfont2=\sevensy \scriptscriptfont2=\fivesy
\textfont\itfam=\tenit \def\it{\fam\itfam\tenit}\def\footnotefont{\ninepoint}%
\textfont\bffam=\tenbf
\def\bf{\fam\bffam\tenbf}\def\sl{\fam\slfam\tensl}\rm}
\font\ninerm=cmr9 \font\sixrm=cmr6 \font\ninei=cmmi9 \font\sixi=cmmi6
\font\ninesy=cmsy9 \font\sixsy=cmsy6 \font\ninebf=cmbx9 \font\nineit=cmti9
\font\ninesl=cmsl9 \skewchar\ninei='177 \skewchar\sixi='177
\skewchar\ninesy='60 \skewchar\sixsy='60
\def\ninepoint{\def\rm{\fam0\ninerm}
\textfont0=\ninerm \scriptfont0=\sixrm \scriptscriptfont0=\fiverm
\textfont1=\ninei \scriptfont1=\sixi \scriptscriptfont1=\fivei
\textfont2=\ninesy \scriptfont2=\sixsy \scriptscriptfont2=\fivesy
\textfont\itfam=\ninei \def\it{\fam\itfam\nineit}\def\sl{\fam\slfam\ninesl}%
\textfont\bffam=\ninebf \def\bf{\fam\bffam\ninebf}\rm}
%
%

\hyphenation{anom-aly anom-alies coun-ter-term coun-ter-terms}
\def\inv{^{\raise.15ex\hbox{${\scriptscriptstyle -}$}\kern-.05em 1}}

\def\Dsl{\,\raise.15ex\hbox{/}\mkern-13.5mu D} 
\def\dsl{\raise.15ex\hbox{/}\kern-.57em\partial}

\font\bigit=cmti10 scaled \magstep1
\def\lspace{\ifx\answ\bigans{}\else\qquad\fi}
\def\lbspace{\ifx\answ\bigans{}\else\hskip-.2in\fi} 
\def\boxeqn#1{\vcenter{\vbox{\hrule\hbox{\vrule\kern3pt\vbox{\kern3pt
     \hbox{${\displaystyle #1}$}\kern3pt}\kern3pt\vrule}
    }}}
\def\mbox#1#2{\vcenter{\hrule \hbox{\vrule height#2in
         \kern#1in \vrule} \hrule}}  


\newwrite\ffile\global\newcount\figno \global\figno=1
\def\nfig#1{\xdef#1{fig.~\the\figno}%
\writedef{#1\leftbracket fig.\noexpand~\the\figno}%
\ifnum\figno=1\immediate\openout\ffile=figs.tmp\fi\chardef\wfile=\ffile%
\immediate\write\ffile{\noexpand\medskip\noexpand\item{Fig.\ \the\figno. }
\reflabeL{#1\hskip.55in}\pctsign}\global\advance\figno by1\findarg}
\def\xfig{\expandafter\xf@g}
\def\xf@g fig.\penalty\@M\ {}
\def\figs#1{figs.~\f@gs #1{\hbox{}}}
\def\f@gs#1{\edef\next{#1}\ifx\next\em@rk\def\next{}\else
\ifx\next#1\xfig #1\else#1\fi\let\next=\f@gs\fi\next}
\newwrite\lfile
{\escapechar-1\xdef\pctsign{\string\%}\xdef\leftbracket{\string\{}
\xdef\rightbracket{\string\}}\xdef\numbersign{\string\#}}

\def\writestop{\def\writestoppt{\immediate\write\lfile{\string\pageno%
\the\pageno\string\startrefs\leftbracket\the\refno\rightbracket%
\string\def\string\secsym\leftbracket\secsym\rightbracket%
\string\secno\the\secno\string\meqno\the\meqno}\immediate\closeout\lfile}}
\def\writestoppt{}\def\writedef#1{}
\def\seclab#1{\xdef #1{\the\secno}\writedef{#1\leftbracket#1}\wrlabeL{#1=#1}}
\def\subseclab#1{\xdef #1{\secsym\the\subsecno}%
\writedef{#1\leftbracket#1}\wrlabeL{#1=#1}}
\newwrite\tfile \def\writetoca#1{}
\def\leaderfill{\leaders\hbox to 1em{\hss.\hss}\hfill}


\def\bar{\overline}
\def\hat{\widehat}

\def\cech{${\rm C}^{\kern-6pt \vbox{\hbox{$\scriptscriptstyle\vee$}\kern2.5pt}}${\rm ech}}
\def\Cech{${\sl C}^{\kern-6pt \vbox{\hbox{$\scriptscriptstyle\vee$}\kern2.5pt}}${\sl ech}}
\def\bfcech{${\bf C}^{\kern-6pt \vbox{\hbox{$\scriptscriptstyle\vee$}\kern2.5pt}}${\bf ech}}


\def\a{{\alpha}}

\def\b{{\beta}}

\def\d{{\delta}}
\def\g{{\gamma}}

\def\e{{\epsilon}}

\def\u{{\Upsilon}}
\def\l{{\lambda}}
\def\s{{\sigma}}
\def\t{{\theta}}


\def\CD{{\cal D}}

\def\CM{{\cal M}}
\def\CN{{\cal N}}
\def\CO{{\cal O}}
\def\CP{{\cal P}}

\def\CT{{\cal T}}
\def\CU{{\cal U}}


\def\bH{{\bf H}}

\def\p{\partial}
\def\pb{\bar{\partial}}

\def\dd{{\rm d}}

\def\ib{\bar{i}}

\def\lb{\bar{l}}

\def\tb{\bar{t}}


%
%

\hyphenation{anom-aly anom-alies coun-ter-term coun-ter-terms}
\def\inv{^{\raise.15ex\hbox{${\scriptscriptstyle -}$}\kern-.05em 1}}

\def\Dsl{\,\raise.15ex\hbox{/}\mkern-13.5mu D} 
\def\dsl{\raise.15ex\hbox{/}\kern-.57em\partial}

\font\bigit=cmti10 scaled \magstep1

\def\lspace{\ifx\answ\bigans{}\else\qquad\fi}
\def\lbspace{\ifx\answ\bigans{}\else\hskip-.2in\fi} 
\def\boxeqn#1{\vcenter{\vbox{\hrule\hbox{\vrule\kern3pt\vbox{\kern3pt
      \hbox{${\displaystyle #1}$}\kern3pt}\kern3pt\vrule}\hrule}}}
\def\mbox#1#2{\vcenter{\hrule \hbox{\vrule height#2in
          \kern#1in \vrule} \hrule}}  
%

\def\darr#1{\raise1.5ex\hbox{$\leftrightarrow$}\mkern-16.5mu #1}

\def\half{{\textstyle{1\over2}}} 
\def\roughly#1{\raise.3ex\hbox{$#1$\kern-.75em\lower1ex\hbox{$\sim$}}}

\def\np#1#2#3{Nucl. Phys. {\bf B#1} (#2) #3}
\def\pl#1#2#3{Phys. Lett. {\bf #1B} (#2) #3}

\def\lmp#1#2#3{Lett. Math. Phys. {\bf #1} (#2) #3}
\def\cmp#1#2#3{Comm. Math. Phys. {\bf #1} (#2) #3}

\def\jhep#1#2#3{JHEP {\bf#1}(#2) #3}

\def\cqg#1#2#3{Class.~Quantum Grav. {\bf #1} (#2) #3}

\def\IB{\relax\hbox{$\inbar\kern-.3em{\rm B}$}}

\def\ID{\relax\hbox{$\inbar\kern-.3em{\rm D}$}}
\def\IE{\relax\hbox{$\inbar\kern-.3em{\rm E}$}}
\def\IF{\relax\hbox{$\inbar\kern-.3em{\rm F}$}}
\def\IG{\relax\hbox{$\inbar\kern-.3em{\rm G}$}}
\def\IGa{\relax\hbox{${\rm I}\kern-.18em\Gamma$}}
\def\IH{\relax{\rm I\kern-.18em H}}
\def\IK{\relax{\rm I\kern-.18em K}}
\def\IL{\relax{\rm I\kern-.18em L}}
\def\IP{\relax{\rm I\kern-.18em P}}
\def\II{\relax{\rm I\kern-.18em I}}


\def\CD{{\cal D}}

\def\CM{{\cal M}}
\def\CN{{\cal N}}
\def\CO{{\cal O}}
\def\CP{{\cal P}}

\def\CT{{\cal T}}
\def\CU{{\cal U}}

\def\bH{{\bf H}}

\def\p{\partial}
\def\pb{\bar{\partial}}

\def\dd{{\rm d}}
\def\ib{\bar{i}}

\def\lb{\bar{l}}

\def\tb{\bar{t}}



\def\inbar{\,\vrule height1.5ex width.4pt depth0pt}

\def\boxit#1{\vbox{\hrule\hbox{\vrule\kern8pt
\vbox{\hbox{\kern8pt}\hbox{\vbox{#1}}\hbox{\kern8pt}}
\kern8pt\vrule}\hrule}}
\def\mathboxit#1{\vbox{\hrule\hbox{\vrule\kern8pt\vbox{\kern8pt
\hbox{$\displaystyle #1$}\kern8pt}\kern8pt\vrule}\hrule}}

\def\lime{{\rm Lim}_{\kern -16pt \vbox{\kern6pt\hbox{$\scriptstyle{\e \to 0}$}}}}

\def\naiveq{\qquad =^{\kern-12pt \vbox{\hbox{$\scriptscriptstyle{\rm naive}$}\kern5pt}} \qquad}

\def\lref{\begingroup\obeylines\lr@f}
\def\lr@f#1#2{\gdef#1{\ref#1{#2}}\endgroup\unskip}


\def\a{{\alpha}}

\def\l{{\lambda}}
\def\lb{{\overline\lambda}}
\def\b{{\beta}}

\def\g{{\gamma}}

\def\d{{\delta}}
\def\e{{\epsilon}}
\def\s{{\sigma}}
\def\k{{\kappa}}

\def\half{{1\over 2}}
\def\p{{\partial}}
\def\pb{{\overline\partial}}
\def\t{{\theta}}
\def\tb{{\overline\theta}}
\def\bar{\overline}

\def\cech{${\rm C}^{\kern-6pt \vbox{\hbox{$\scriptscriptstyle\vee$}\kern2.5pt}}${\rm ech}}
\def\Cech{${\sl C}^{\kern-6pt \vbox{\hbox{$\scriptscriptstyle\vee$}\kern2.5pt}}${\sl ech}}

\Title{\vbox{\hbox{IFT-P.030/2006 }\hbox{IHES-P/06/41}\hbox{ITEP-TH-39/06}}}
{\vbox{
\medskip
\centerline{MULTILOOP SUPERSTRING AMPLITUDES}
\medskip
\centerline{FROM NON-MINIMAL}
\medskip
\centerline{PURE SPINOR FORMALISM}}}
\bigskip\centerline{Nathan Berkovits\foot{\tt e-mail: nberkovi@ift.unesp.br}
and Nikita Nekrasov\foot{\tt e-mail: nikita@ihes.fr}\footnote{}{On leave of 
absence from: ITEP, Moscow, Russia}}
\bigskip
\centerline{$^{1}$ \it Instituto de F\'\i sica Te\'orica, State University of
S\~ao Paulo}
\centerline{\it Rua Pamplona 145, 01405-900, S\~ao Paulo, SP  Brasil}
\centerline{$^{2}$ \it Institut des Hautes Etudes Scientifiques}
\centerline{\it Le Bois-Marie, 35 route de Chartres, 91440 Bures-sur-Yvette France}
\vskip .3in
\noindent
{}Using the non-minimal version of the pure spinor formalism, 
manifestly super-Poincar\'e covariant superstring
scattering amplitudes can be computed as in topological string
theory without the need of picture-changing operators.
The only subtlety comes from 
regularizing the functional integral over the pure spinor ghosts.
In this paper, it is shown how to regularize this functional integral in a 
BRST-invariant manner, allowing the computation of arbitrary multiloop
amplitudes. The regularization method simplifies for
scattering amplitudes which contribute to ten-dimensional
F-terms, i.e. terms in the ten-dimensional
superspace action which do not involve integration over the maximum
number of $\theta$'s.

\vskip .3in

\Date{August 2006}


\lref\natnik{N.~Berkovits, N.~Nekrasov, {\it The character of pure spinors}, hep-th/0503075}
\lref\cartan{
E. Cartan, {\it The Theory of Spinors}, Dover, New York, 1981\semi P. Budinich and
A. Trautman, {\it The Spinorial Chessboard},  Trieste Notes in Physics,
Springer-Verlag, Berlin, 1988\semi
C.~Chevalley, {\it The algebraic theory of spinors}, Columbia University Press, NY 1954}
\lref\berkovitsbghost{N.~Berkovits, 
{\it Multiloop amplitudes and vanishing theorems using the pure spinor formalism for the superstring}, \jhep{0409}{2004}{047}, hep-th/0406055\semi
N.~Berkovits, {\it Covariant Multiloop Superstring Amplitudes}, hep-th/0410079}
\lref\berkovitsmulti{N.~Berkovits, {\it  Multiloop Amplitudes and Vanishing Theorems using the Pure Spinor Formalism for the Superstring}, hep-th/0406055}
\lref\natser{N.~Berkovits, S.~Cherkis, {\it Higher dimensional twistor transforms using pure spinors},
hep-th/0409243}
\lref\pst{N.~Berkovits, {\it Pure Spinor Formalism as an ${\CN}=2$ Topological String}, \jhep{0510}{2005}{089}, hep-th/0509120}
\lref\berkovitstrieste{N.~Berkovits, {\it ICTP Lectures on covariant quantization of the superstring}, 
hep-th/0209059}
\lref\howe{G.~Moore, P.~Nelson, {\it The aetiology of sigma model anomalies}, \cmp{100}{1985}{83-132}\semi
P.~S.~Howe, G.~Papadopoulos, {\it Anomalies in two-dimensional supersymmetric non-linear $\s$-models}, \cqg{4}{1987}{1749-1766}}
\lref\gravan{L.~Alvarez-Gaume, E.~Witten, {\it Gravitational anomalies}, \np{234}{1983}{269-330}}
\lref\gsw{M.~B.~Green, J.~H.~Schwarz, {\it Covariant description of superstrings}, \pl{136}{1984}{367-370}}
\lref\gsan{M.~B.~Green, J.~H.~Schwarz, {\it Anomaly cancellations in supersymmetric $d=10$ gauge theory and superstring theory}, \pl{149}{1984}{117-122}}
\lref\witohet{E.~Witten, {\it Some properties of $O(32)$ superstrings}, \pl{149}{1984}{351-356}} 
\lref\technique{A.~Schwarz, hep-th/0102182}
\lref\membrane{N. Berkovits, {\it Toward Covariant Quantization of the Supermembrane}, 
\jhep{0209}{2002}{051}, hep-th/0201151.}
\lref\vanhove{L. Anguelova, P.A. Grassi and P. Vanhove, {\it
Covariant one-loop amplitudes in D=11}, Nucl. Phys. B702 (2004) 269,
hep-th/0408171\semi P.A. Grassi and P. Vanhove, {\it
Topological M Theory from Pure Spinor Formalism}, hep-th/0411167.}
\lref\golos{A.~Gorodentsev and A.~Losev, around late 2000, unpublished.}
\lref\lmt{A.~Losev, A.~Marshakov, A.~Tseitlin, {\it On first order formalism in string theory}, hep-th/0510065}
\lref\movsch{M.~Movshev and A.~Schwarz,"On maximally supersymmetric
Yang-Mills theories", Nucl. Phys. B681 (2004) 324, hep-th/0311132\semi
M.~Movshev and A.~Schwarz, "Algebraic structure of Yang-Mills
theory", hep-th/0404183.}
\lref\abcd{N.~Nekrasov and S.~Schadchine, {\it ABCD of instantons},
\cmp{252}{2004}{359-391}, hep-th/0404225.}
\lref\FS{L.~D.~Faddeev, S.~L.~Shatashvili, {\it Realization of the Schwinger term in the Gauss law and the possibility of correct quantization of a theory with anomalies}, \pl{167}{1986}{225-228}\semi
L.~D.~Faddeev, S.~L.~Shatashvili, {\it Algebraic and Hamiltonian methods in the theory of non-abelian anomalies}, Theor. Math. Phys. {\bf 60} (1984) 770}
\lref\atiyahsinger{M.~F.~Atiyah, I.~M.~Singer, {\it Dirac operators coupled to vector bundles}, Proc. Natl. Acad. Sciences, {\bf 81} (1984) 2597}
\lref\atiyahbott{M.~Atiyah, R.~Bott, {\it The moment map and equivariant cohomology} , Topology  {\bf  23}, vol. 1(1984) 1-28}
\lref\samsonco{A.~Alekseev, L.~Fadeev, S.~Shatashvili, {\it Quantization of symplectic orbits of compact Lie groups by means of the functional integral}, J.of Geometry and Physics, {\bf 5} (3) (1988) 391-406}
\lref\kliu{K.~Liu, {\it Holomorphic equivariant cohomology}, Math. Annalen, {\bf 303} 1 ( 1995) 125-148}
\lref\pietropeter{P.~A.~Grassi, G.~Policastro , M.~Porrati,  P.~Van~Nieuwenhuizen, {\it Covariant quantization of superstrings without pure spinor constraints}, \jhep{0210}{2002}{054},
hep-th/0112162\semi Y. Aisaka and Y. Kazama, {\it A new first class
algebra, homological perturbation and extension of
pure spinor formalism for superstring}, \jhep{0302}{2003}{017}, hep-th/0212316}
\lref\opennc{N.~Nekrasov, {\it Lectures on open strings, and noncommutative gauge fields}, hep-th/0203109}
\lref\cdv{
A.~Connes, M.~Dubois-Violette, {\it Yang-Mills algebra}, math.QA/0206205, \lmp{61}{2002}{149-158} }
\lref\feiginfrenkel{B.~Feigin, E.~Frenkel, {\it Affine Kac-Moody Algebras and Semi-Infinite Flag Manifolds}, \cmp{128}{1990}{161-189} }
\lref\ff{B.~Fegin, E.~Frenkel, {\it A family of representations of affine Lie algebras}, Rus. Math. Surveys {\bf 43} (1988), 221-222\semi
{\it Representations of affine Kac-Moody algebras and bosonization}, in {\sl Physics and mathematics of strings}, pp. 271-316, World Scientific, 1990\semi
{\it Representations of affine Kac-Moody algebras, bosonoziation and resolutions}, Lett. Math. Phys.{\bf 19} (1990) 307-317}

\lref\edik{E.~Frenkel, {\it Wakimoto modules, opers and the center at the critical level}, Adv. in Math. {\bf 195} (2005)  197-404, math.QA/0210029}
\lref\jan{J.~de Boer, L.~Feher, {\it Wakimoto realizations of current algebras: an explicit construction}, \cmp{189}{1997}{759-793}}
\lref\bln{A.~Losev, L.~Baulieu, N.~Nekrasov, {\it Target space symmetris in topological field theories I}, \jhep{0202}{2002}{021}, hep-th/0106042}
\lref\kapustin{ F.~Malikov, V.~Schekhtman, A.~Vaintrob, "Chiral de Rham complex",  $\qquad\qquad$ math.AG/9803041\semi
F.~Malikov, V.~Schekhtman, "Chiral Poincare duality", math.AG/9905008\semi
F.~Malikov, V.~Schekhtman, "Chiral de Rham complex II", math.AG/9901065\semi
V.~Gorbounov, F.~Malikov, V.~Schekhtman,  "Deformations of chiral algebras and quantum cohomology of toric varieties", math.AG/0001170\semi
E.~Frenkel, M.~Szczesny, {\it Chiral de Rham complex and orbifolds}, math.ag/0307181\semi
A.~Kapustin, {\it Chiral de Rham complex and the half-twisted sigma model}, hep-th/0504074}
\lref\ztwitten{E~Witten, {\it Two dimensional models with (0,2) supersymmetry: perturbative aspects}, hep-th/0504078}
\lref\alekseev{A.~Alekseev, T.~Strobl, "Current Algebras and Differential Geometry", hep-th/0410183}

\lref\malikov{V.~Gorbounov, F.~Malikov, V.Schekhtman, "On chiral differential operators over homogeneous spaces", math.AG/0008154\semi
V.~Gorbounov, F.~Malikov, V.~Schekhtman,  "Gerbes of chiral differential operators, I-III", math.AG/0005201, math.AG/0003170, math.AG/9906117}
\lref\cov{N. Berkovits, {\it
Super-Poincar\'e Covariant Quantization of the
Superstring}, \jhep{04}{2000}{018}, hep-th/0001035}

\lref\griffitsharris{P.~Griffiths, J.~Harris, {\it Principles of algebraic geometry}, 1978, New York, Wiley \& Sons}

\lref\sharpe{E.~Sharpe, {\it Lectures on D-branes and sheaves}, hep-th/0307245} 
\lref\courant{T.J.~Courant, {\it Dirac manifolds}, Trans. AMS {\bf 319} (1990), 631-661}

\lref\hitchin{N.~Hitchin, {\it Generalized Calabi-Yau manifolds}, math.DG/0209099}
\lref\gualtieri{M.~Gualtieri, {\it Generalized complex geometry}, math.DG/0401221}
\lref\strobl{A.~Kotov, P.~Shaller,  T.~Strobl, {\it Dirac sigma models}, hep-th/0411112}

\lref\nekbg{N.~Nekrasov, {\it Lectures on curved beta-gamma systems, pure spinors, and anomalies}, IHES-P-05-35, ITEP-TH-58-05, hep-th/0511008}
\lref\fms{D.~Friedan, E.~Martinec, S.~Shenker, {\it Covariant quantization of superstrings}, 
\pl{160}{1985}{55} \semi
{\it Conformal invariance, supersymmetry and string theory}, \np{271}{1986}{93}}
\lref\knizhnik{V.~Knizhnik, {\it Analytic fields on Riemann surfaces}, \pl{180}{1986}{247} \semi
{\it Analytic fields on Riemann surfaces II}, \cmp{112}{1987}{567-590}}
\lref\fln{E.~Frenkel, A.~Losev, N.~Nekrasov, {\it Instantons and conformal field theories}, to appear}
\lref\aib{E.~Frenkel, A.~Losev, {\it Mirror symmetry in two steps: A-I-B}, hep-th/0505131}
\lref\swnc{N.~Seiberg, E.~Witten, {\it String theory and noncommutative geometry}, JHEP 09 (1999) 032}
\lref\bcov{M.~Bershadsky, S.~Cecotti, H.~Ooguri, C.~Vafa, {\it Kodaira-Spencer theory of gravity and exact results for quantum string amplitudes}, \cmp{165}{1994}{311-428}, hep-th/9309140}
\lref\bpz{A.~Belavin, A.~Polyakov, A.~Zamolodchikov, {\it Infinite conformal symmetry in two dimensional quantum field theory}, \np{241}{1984}{333-380}}
\lref\polyakov{A.~Polyakov, {\it Quantum geometry of bosonic strings}, \pl{103}{1981}{207-210}\semi
A.~Polyakov, {\it Quantum geometry of fermionic strings}, \pl{103}{1981}{211-213}}
\lref\verlinde{M.~Rocek, E.~Verlinde, {\it Duality, Quotients, and Currents}, hep-th/9110053}
\lref\sammar{E.~Martinec, S.~Shatashvili, {\it Black hole physics and Liouville theory}, \np{368}{1992}{338-358}}
\lref\arkadyalbert{A.~Tseytlin, A.~Schwarz, {\it Dilaton shift under duality and torsion of elliptic complex}, hep-th/9210015}
\lref\wittentdg{E.~Witten, {\it Two dimensional gauge theories revisited}, hep-th/9204083}

\newsec{Introduction}

Over the past six years, a new formalism combining the nice features of
Ramond-Neveu-Schwarz and Green-Schwarz approaches to quantization
of the superstring \cov\ has been developed. The formalism is called the 
pure spinor formalism as it involves twistor-like variables which take 
values in 
the space $\CP$ of pure spinors in ten dimensions. 

More precisely, 
the pure spinor formalism involves a geometric sigma model
describing the maps of the worldsheet $\Sigma$ to ten-dimensional
super-Minkowski space, together with a
somewhat unconventional {\it curved $\b\g$-system} which describes 
the maps of $\Sigma$ to the space $\CP$. 
In the minimal pure spinor formalism,
only the worldsheet fields $\l^\a$ which are holomorphic coordinates
on $\CP$ are used.

Although one can compute scattering amplitudes using the minimal
pure spinor formalism,
the absence of a composite $b$ ghost
in the minimal formalism makes the amplitude prescription non-conventional.
In the approach of 
\berkovitsbghost, 
picture-changing operators are used to
construct a picture-raised version of the $b$ ghost. Unfortunately,
these picture-changing operators are only Lorentz-invariant up to
BRST-trivial terms, so manifest Lorentz covariance is broken at
intermediate stages in the computation.

A more elegant approach \ref\niklec{N.~Nekrasov, {\it 
Lectures at the 23d Jerusalem Winter School in Theoretical Physics}, The Hebrew University Jerusalem, January 2-5, 2006\semi {\tt http://www.as.huji.ac.il/schools/phys23/media2.shtml}} uses the so-called {\cech} 
cohomology language, where the $b$ ghost is viewed as a
collection of {\cech} cochains of various degrees, from zero to three, 
on the space of pure spinors:
\eqn\bcech{b = \left( b_{\a} \right) + \left( b_{\a\b} \right) + 
\left( b_{\a\b\g} \right) + \left( b_{\a\b\g\d} \right) .} 
Here ${\a}$ etc. label the coordinate patches ${\CU}_{\a}$ on the space
of pure spinors:
\eqn\prusn{{\CP} = \cup_{\a} \ {\CU}_{\a}}
and one can choose the coordinate patches to be in one-to-one
correspondence with the components of an unconstrained spinor, i.e.
${\a} = 1, \ldots , 16$.  On the coordinate patch ${\CU}_{\a}$,
the component ${\l}^{\a}$ of the pure spinor is not allowed to vanish.

In this approach, manifest Lorentz invariance is broken by working
with 
{\cech} 
cochains which are defined on 
the (intersections of) coordinate patches, in our case on the 
space $\CP$ of pure spinors. Although the space of pure spinors is 
a homogeneous space of the (euclidean version of the) Lorentz group, a 
particular coordinate patch is not. So, it is aesthetically more 
appropriate to work in a formalism where the choices of coordinates 
on $\CP$ are not necessary. 

A well-known alternative to 
the {\cech} language in algebraic geometry is Dolbeault 
language where instead of the locally defined 
holomorphic objects, one deals with the globally defined 
non-holomorphic ones. 
In the context of two dimensional sigma models, the 
Dolbeault version is formulated by including the anti-holomorphic 
coordinates ${\lb}_{\a}$ on $\CP$ and the fermionic coordinates 
$r_{\a} = {\dd}{\lb}_{\a}$. In principle,
there are two options -- one can treat these new coordinates as fields
of the same worldsheet chirality as ${\l}^{\a}$, as will be done
in this paper following 
\pst, or as fields of 
the opposite worldsheet chirality, as is more natural in the
context of $(0,2)$ 
models \ztwitten.

After including the non-minimal worldsheet fields $(\lb_\a, r_\a)$,
it was shown in \pst\
how to
construct a composite $b$ ghost and compute superstring scattering
amplitudes as in topological string theory
without picture-changing insertions. 
The only subtletly in this non-minimal prescription is that
the composite $b$ ghost contains factors of $(\l^\a \lb_\a)^{-1}$.
When functionally integrating over $\l^\a$ and $\lb_\a$
in the scattering amplitude,
these $(\l^\a \lb_\a)^{-1}$ factors can cause a problem coming from
the functional integration region
where all components of $\l^\a$ are zero. In \pst, it was shown that
this problem can be avoided for amplitude computations up to two loops, 
but it was not shown how to resolve this problem for computations with
more than two loops.

In this paper, it will be shown how to resolve this problem for
arbitrary multiloop amplitudes by 
constructing a regularized version of the $b$ ghost, $b_\e$,
which is non-singular when $(\l^\a \lb_\a) \to 0$.
The regularized $b$ ghost will be defined as 
\eqn\Mreg{ b_\e = e^{-\e (w_\a \bar w^\a + ...)} b}
where $\e$ is a positive constant,
$w_\a$ and $\bar w^\a$ are the conjugate momenta to $\l^\a$ and
$\lb_\a$, and $...$ is chosen such that $(w_\a \bar w^\a + ...)$ is
well-defined, i.e. gauge invariant, and BRST-trivial.

This regularization procedure can be viewed as an analogue of turning
on a metric perturbation in the $(0,2)$ model. Indeed, 
 $w_\a \bar w^\a$ acts essentially as the Laplacian on functions
of pure spinors. If $\bar w^{\a}$ and $w_{\a}$ were fields of opposite
worldsheet chirality, the term $w_{\a} {\bar w}^{\a}$ would serve as
an inverse metric perturbation of a curved beta-gamma system:
\eqn\bgpert{{\b}_{i} {\pb} {\g}^{i} + {\bar\b}_{\ib} {\p} {\bar \g}^{\ib} 
+ {\a}^{\prime} g^{i\ib} {\b}_{i} {\bar \b}_{\ib}}
which preserves
conformal invariance for special metrics that are Ricci-flat
(in the first order approximation) \lmt. In our case where
the fields $w_{\a}$ and
${\bar w}^{\a}$  have the same chirality, perturbations
like \bgpert\ would break conformal invariance. 
Nevertheless, both in the context of $(0,2)$ models and in our case,
perturbations like \bgpert\ can be made $Q$-exact. Thus, 
conformal invariance would  be preserved at the level of $Q$-cohomology. 

Also, let us mention the r{\^o}le of metric perturbations \bgpert\
in the context of conventional topological strings 
obtained by twisting $(2,2)$ supersymmetric sigma models. 
In particular, for the
A twist one obtains the theory with the Lagrangian \aib:
\eqn\topstr{{\b}_{i} {\pb} {\g}^{i} + {\bar\b}_{\ib} {\p} {\bar \g}^{\ib} + b_{i} {\pb} c^{i} + {\bar b}_{\ib} {\p} {\bar c}^{\ib}}  
that ensures that the path integral localizes onto the space
of holomorphic maps of the worldsheet into the complex target space, which
has an infinite radius metric in this description. 
This theory is well-defined (for compact targets) on genus zero,
but for higher genera ( in the trivial instanton sector) the 
zero modes of the ${\b}, {\bar \b} , b , {\bar b}$
fields need regularization. One can turn on the deformation
to finite radius by adding the term:
\eqn\finrad{  
{\a}^{\prime} \{ Q , {\half} g^{i\ib} 
( b_{i} {\bar \b}_{\ib} + {\bar b}_{\ib} {\b}_{i} ) \} = 
{\a}^{\prime} \left( g^{i\ib} {\b}_{i} {\bar \b}_{\ib} 
+ {\rm fermions} \right) }
which, upon proper treatment of the coupling to worldsheet 
topological gravity, induces
the celebrated \bcov\ $ c \left( {\bH}_{g} \otimes {\CT}_{X}) \right) $
measure on the moduli space of Riemann surfaces of genus $g$.  

Since \Mreg\ will be defined such that
$b_\e = b +  \{Q,\Omega_\e\}$ for some $\Omega_\e$, 
BRST-invariant amplitudes are unaffected by the replacement of $b$ with
$b_\e$ in the amplitude prescription of \pst. 
Using the regularized $b$ ghost, the 
multiloop amplitude prescription in the non-minimal
pure spinor formalism is therefore given by
\eqn\regampone{A = \lim_{\e\to 0}\int d^{3g-3 }\tau \
\Biggl\langle \prod_{r=1}^N \int dz_r U_r(z_r) \prod_{s=1}^{3g-3} 
\int (\mu_s b_\e)  ~{\cal N}
\Biggr\rangle }
where
$\int d^{3g-3 }\tau
\langle \prod_{r=1}^N \int dz_r U_r(z_r) \prod_{s=1}^{3g-3} \int (\mu_s b) 
\rangle$ is the usual prescription of bosonic string theory, 
${\cal N}$ is the zero mode normalization factor defined in \pst\
which regularizes the functional integral over the zero modes, and
the right-moving contribution to $A$ is being ignored.
Since BRST invariance implies
that \regampone\ is independent of the parameter $\e$, 
one can take the limit $\e\to 0$
at the end of the computation.

The choice of the $\e$-regularization is not unique. In the present paper
we give one such choice, in order to demonstrate that a
completely regular expression for the amplitudes exists. 
Although our construction can be motivated by the
considerations of \finrad, there
may well exist a much simpler regularization. At present, 
for generic multiloop superstring amplitude computations, it is
still difficult to evaluate the limit $\e\to 0$ of \regampone\ since our 
expression for $b_{\e}$ is rather complicated.

However, for certain special amplitudes
in which not all $\t^\a$ zero modes are absorbed by the external 
vertex operators, the $\e$-regularization of the $b$ ghost
is unnecessary,
and it is easy to take the limit
$\e\to 0$ of \regampone. 
These special amplitudes can contribute to ten-dimensional F-terms in the
effective action, i.e. terms in which
the superspace action involves integration over fewer than 16 $\t$'s
for N=1 D=10, or fewer than 32 $\t$'s for N=2 D=10.

It is interesting that topological string methods are also useful for
computing F-terms in the four-dimensional effective action coming from
Calabi-Yau compactification \ref\anton{I. Antoniadis, E. Gava, K.S. Narain
and T.R. Taylor, {\it Topological amplitudes in string theory},
Nucl. Phys. B413 (1994) 162, hep-th/9307158.}\bcov,\ref\hyb{N. Berkovits
and C. Vafa, {\it ${\CN}=4$ Topological Strings}, Nucl. Phys. B433 (1995) 123,
hep-th/9407190.}.
In \pst, it was shown that these lower
dimensional F-term computations can be reproduced using a compactified
four-dimensional version of the pure spinor formalism.

The paper is organized as follows. In section $\underline{2}$, the non-minimal
pure spinor formalism will be reviewed. In section $\underline{3}$, the 
regularized
$b$ ghost will be constructed using the heat kernel method.
In section $\underline{4}$, 
the multiloop amplitude prescription will be defined using the
regularized $b$ ghost $b_{\e}$,
and it will be shown that this prescription simplifies
for amplitudes which contribute
to ten-dimensional F-terms.

\newsec{Review of Non-Minimal Formalism}

\subsec{Minimal formalism}

The minimal pure spinor
formalism for the superstring is constructed using the $(x^m,\t^\a)$
variables of $d=10$ superspace where $m=0$ to 9 and $\a=1$ to 16,
together with the fermionic conjugate momenta $p_\a$. Furthermore, one
introduces a bosonic spinor ghost $\l^\a$
which satisfies the pure spinor constraint
\eqn\puredef{\l^\a \g^m_{\a\b}\l^\b=0}
where $\g^m_{\a\b}$ are the symmetric $16\times 16$ $d=10$ Pauli matrices.
Because of the pure spinor constraint on $\l^\a$, its
conjugate momentum $w_\a$ is defined up
to the gauge transformation
\eqn\gaugeone{\d w_\a = \Lambda^m (\g_m\l)_\a,}
which implies that $w_\a$ only appears through its Lorentz current
$N_{mn}$, ghost current $J$, and
stress tensor $T_\l$. These gauge-invariant currents are defined by
\eqn\currone{N_{mn}=\half w\g_{mn}\l,\quad J = w_\a\l^\a,\quad
T_\l = w_\a \p\l^\a.}

The worldsheet action for the left-moving matter and ghost variables is
\eqn\actionpure{S=\int d^2 z \ \left( \half \p x^m \bar\p x_m + p_\a \bar\p\t^\a
- w_\a \bar\p\l^\a \right) \ ,}
and the right-moving variables will be ignored throughout this paper.
For the Type II superstring, the right-moving variables are similar to
the left-moving variables, while for the heterotic superstring, the
right-moving variables are the same as in the RNS heterotic formalism.

Physical open string states in the minimal pure spinor formalism are defined
as ghost-number one states in the cohomology of the nilpotent BRST
operator
\eqn\purebrst{Q=\int dz ~\l^\a d_\a}
where
\eqn\ddef{d_\a= p_\a -\half\g_{\a\b}^m \t^\b \p x_m -{1\over 8}
\g_{\a\b}^m \g_{m\g\d}\t^\b\t^\g\p\t^\d}
is the supersymmetric Green-Schwarz constraint.

Although one can compute scattering amplitudes using the minimal formalism,
the lack of a composite
$b$ ghost satisfying $\{Q,b\}=T$ makes
the amplitude prescription unconventional. It is easy to
see that the minimal formalism does not contain such a composite
$b$ ghost since the gauge-invariant combinations of $w_\a$ in
\currone\ all carry zero ghost number, so there are no gauge-invariant
operators of negative ghost number.

\subsec{Non-minimal worldsheet variables}

As shown in \pst, this difficulty can be resolved by adding non-minimal
variables to the formalism which allow the construction
of a composite $b$ ghost.
The new non-minimal variables consist of a bosonic pure spinor
$\bar\lambda_\a$ and
a constrained fermionic spinor $r_\a$ satisfying the constraints
\eqn\newcons{\lb_\a\g_m^{\a\b}\lb_\b=0 \quad {\rm and}\quad
\lb_\a \g_m^{\a\b} r_\b=0.}
In d=10 Euclidean space where complex conjugation flips the chirality
of spacetime spinors, $\lb_\a$ can be interpreted as the complex conjugate
to $\l^\a$.
The worldsheet action for the non-minimal pure spinor formalism is
\eqn\nonmin{\int d^2 z \ \left( \half \p x^m \bar\p x_m +p_\a \bar\p\t^\a
- w_\a\bar\p \l^\a - \bar w^\a \bar\p \bar\l_\a + s^\a \bar\p r_\a \right)}
where $\bar w^\a$ and $s^\a$ are the conjugate momenta for $\bar\l_\a$
and $r_\a$ with $+1$ conformal weight.

Just as $w_\a$ can only appear in the gauge-invariant combinations
\eqn\mingauge{
N_{mn}= \half (w\g_{mn}\l),\quad J=w_\a\l^\a,\quad T_\l = w_\a\p\l^\a,}
the variables $\bar w^\a$ and $s^\a$ can only appear in the combinations
\eqn\gaugeinvbar{\bar N_{mn}= \half (\bar w\g_{mn}\lb - s\g_{mn} r),\quad
\bar J =\bar w^\a \bar\l_\a -s^\a r_\a,\quad
T_\lb = \bar w^\a\p\bar\l_\a - s^\a \p r_\a,}
$$S_{mn} = \half s \g_{mn} \lb,\quad  S=  s^\a \lb_\a,$$
which are invariant under the gauge transformations
\eqn\gaugebar{\d \bar w^\a = \bar\Lambda^m (\g_m\lb)^\a - \phi^m (\g_m r)^{\a}
,
\quad \d s^\a = \phi^m (\g_m \lb)^\a}
for arbitrary $\bar\Lambda^m$ and $\phi^m$.

{}In order that the non-minimal variables do not affect the cohomology,
the ``minimal'' pure spinor BRST operator $Q=\int dz\  \l^\a d_\a$ will
be modified to the ``non-minimal'' BRST operator 
\eqn\nonminQ{Q_{nonmin} = \int dz \ \left( \l^\a d_\a + \bar w^\a r_\a \right)\ .}
The new term $\int dz \bar w^\a r_\a$ is invariant under the gauge
transformation of \gaugebar\
and implies through the usual quartet argument that
the cohomology is independent of $(\bar \l_\a, \bar w^\a)$ and
$(r_\a, s^\a)$.

{}The ghost-number operator in the non-minimal formalism is naturally
defined as 
\eqn\ghnnm{\int dz (\l^\a w_\a - \lb_\a \bar w^\a) }
so that
$\l^\a$ carries ghost-number $+1$ and $\lb_\a$ carries ghost-number
$-1$. The corresponding ghost-number anomaly was computed in \pst\ to
be $+3$, so the non-minimal formalism can be treated as a 
critical topological string theory.

{}A simple way to understand the value $+3$ of the ghost number anomaly is to look at the way the measure for the non-minimal fields is defined. 
The issue is the zero modes. On a genus $g$ Riemann surface, the field
$w_{\a}$ has $11 g$ zero modes, ${\l}^{\a}$ has $11$ zero modes,
and similarly for ${\bar w}^{\a}, {\bar \l}_{\a}, s^{\a}$ and $r_{\a}$.
The measure on $ w, \l$ zero modes is defined using the holomorphic top 
form $\Omega$ on the space $\CP$ of pure spinors:
\eqn\wlms{{\CD} w_{\a} {\CD} {\l}^{\a} \sim {\Omega}^{1-g} \ ,\
{\Omega} = {{\dd}^{11}{\l} \over {\l}^{3}} = {\l}_{+}^{7} 
{\dd} {\l}_{+} \wedge {\dd}^{10} u_{ab}}
where we used the local parameterization of the pure spinor in the form:
$$
{\l} = {\l}_{+} \left( 1 , u_{ab} , u_{[ab} u_{cd]} \right)
$$
The form $\Omega$ has weight $+8$ under the symmetry generated
by \ghnnm. So, the measure factor ${\Omega}^{1-g}$ has charge $8 (1- g)$ on the
genus $g$ surface. At the same time, the measure on ${\bar\lambda}$, 
$r$ fields is defined canonically, as the fermions $r$ 
are in the antiholomorphic tangent bundle to $\CP$. The 
zero modes of $\bar\lambda$, $r$ and the corresponding momenta bring a factor
$$
{\CD} {\bar w}^{\a} {\CD} {\bar\l}_{\a} 
{\CD} s^{\a} {\CD} r_{\a}  \sim {\bar\Omega}^{1-g}
$$
where
$$
\bar\Omega = {\dd}^{11}{\bar\lambda} {\dd}^{11} r
$$
has charge $-11$ under \ghnnm. Thus, the total anomalous charge 
is $+3 ( g -1)$ on the genus $g$ Riemann surface, as claimed.

\subsec{{\bfcech} and Dolbeault}

The addition of non-minimal variables and the construction of
$Q_{nonmin}$ can be understood as standard techniques which
are used in relating 
{\cech} and Dolbeault cochains. 
To describe {\cech} cochains, first express the space of pure spinors
$\CP$ as the union of coordinate patches $U_\a$ for $\a=1$ to 16
where the component $\l^\a$ of the pure spinor is required to
be non-vanishing on $U_\a$.  
The analysis of anomalies of the curved $\b\g$-system on the 
pure spinor space 
implies that the point ${\l} = 0$ is not in $\CP$
\nekbg, thus one can
always find $\a$ such that a given point ${\l} \in \CP$ belongs to the
coordinate patch ${\CU}_{\a}$. So 
$\CP = \cup_{\a} {\CU}_{\a}$.

The {\cech} $k$-cochain is an object
${\psi}_{{\a}_{0} {\a}_{1} \ldots {\a}_{k}}$ 
which is holomorphic 
on the intersection
\eqn\intr{{\CU}_{{\a}_{0} {\a}_{1} \ldots {\a}_{k}} 
= {\CU}_{{\a}_{0}} \cap {\CU}_{{\a}_{1}} \cap \ldots \cap {\CU}_{{\a}_{k}}}
and which obeys
\eqn\chckdl{( {\d}{\psi})_{{\a}_{0}{\a}_{1} \ldots {\a}_{k}} \equiv
{\psi}_{{\a}_{1}{\a}_{2}\ldots {\a}_{k}} - 
{\psi}_{{\a}_{0}{\a}_{2}\ldots {\a}_{k}} + \ldots (-1)^{k} 
{\psi}_{{\a}_{0}{\a}_{1}\ldots {\a}_{k-1}}  = 0 \ .}
The standard way to relate {\cech} and Dolbeault cochains is to use
the so-called {\it partition of unity} (cf. \ztwitten).  
In the case of the space of pure spinors it can be taken to be:
\eqn\partun{{\rho}_{\a} = {1\over {(\bar\l}{\l})} {\bar\l}_{\a} {\l}^{\a} }
where
the functions ${\rho}_{\a}$ vanish outside the corresponding domains
${\CU}_{\a}$, and they sum to unity
$$
\sum_{{\a} = 1}^{16} {\rho}_{\a} = 1.
$$
Note that $(\lb \l)$ denotes
$\sum_{\b=1}^{16} \lb_\b \l^\b$ and repeated indices are
not assumed to be summed over in this subsection.

{}Now, given a {\cech} cochain satisfying \chckdl,
one can define the corresponding Dolbeault cocycle, i.e. a ${\pb}$-closed
differential form of type $(0,p-1)$:
\eqn\dlbfrm{{\hat\psi} = {1\over p!} \sum_{{\a}_{1} ,   \ldots ,   {\a}_{p}}
{\psi}_{{\a}_{1}\ldots {\a}_{p}} {\rho}_{{\a}_{1}}
{\pb} {\rho}_{{\a}_{2}} \wedge \ldots \wedge {\pb} {\rho}_{{\a}_{p}}.}
An important generalization of this well-known construction consists of
replacing the function valued cochains $({\psi}_{\ldots})$ 
by cochains which take values in some (super)commutative 
algebra, of even complex, and by replacing
the operator ${\pb}$ by the general operator
${\pb} + Q$ where $Q$ is the differential in this complex. 
The resulting globally defined form ${\hat\psi}$ obeys
$$
( {\pb} + Q ) {\hat \psi}  = 0
$$
iff the {\cech} cochain verifies 
$$
( {\d} + Q ) {\psi} = 0.
$$
To compare with the non-minimal formalism described in the previous
subsection, define 
\eqn\debq{Q = \sum_{\a=1}^{16} \int dz\ \l^\a d_\a \quad  {\rm and}
\quad
\pb = \sum_{\a=1}^{16} \int dz\ \bar w^\a r_\a}
so that $Q_{nonmin} \hat\psi = (\pb + Q) \hat\psi$.
Using this definition,
one finds in \dlbfrm\ that
\eqn\dbarrho{
\pb \rho_\a = {{(\lb \l) r_\a - (r\l) \lb_\a}\over {(\l\lb)^2}} {\l}^{\a} .}

\subsec{Construction of $b$ ghost}

Although there is no globally defined operator in the minimal formalism
satisfying $\{Q, b\}=T$, a $b$ ghost can be constructed 
in the non-minimal formalism using the operators $[G^\a, H^{[\a\b]},
K^{[\a\b\g]}, L^{[\a\b\g\d]}]$ which carry zero ghost-number 
and satisfy \pst\berkovitsbghost
\ref\tonin{ 
I.~Oda, M.~Tonin, {\it On the b-antighost in the pure spinor 
quantization of superstrings},  \pl{606}{2005}{218-222}, hep-th/0409052\semi
I.~Oda, M.~Tonin, {\it Y-formalism in pure spinor 
quantization of superstrings}, Nucl. Phys. B727 (2005)176,
hep-th/0505277\semi
I.~Oda, M.~Tonin, {\it The b-field in pure spinor 
quantization of superstrings}, hep-th/0510223.
}
\eqn\ghkl{\{Q, G^\a\} = \l^\a T, \quad [Q,H^{[\a\b]}]=\l^{[\a} G^{\b]},\quad
\{Q,K^{[\a\b\g]}\}=\l^{[\a} H^{\b\g]},}
$$
[Q,L^{[\a\b\g\d]}]=\l^{[\a} K^{\b\g\d]},\quad
\l^{[\a} L^{\b\g\d\k]}=0. $$

As discussed in \niklec, this construction can be naturally understood in
{\cech} language by defining 
\eqn\bcech{b = \left( b_{\a} \right) + \left( b_{\a\b} \right) + 
\left( b_{\a\b\g} \right) + \left( b_{\a\b\g\d} \right) } 
where 
$[\left( b_{\a} \right), \left( b_{\a\b} \right), 
\left( b_{\a\b\g} \right),\left( b_{\a\b\g\d} \right) ]$ are {\cech}
cochains of degree zero to three defined by
\eqn \bcoch{(b_\a) = {{G^\a}\over {\l^\a}},\quad
(b_{\a\b}) = {{H^{[\a\b]}}\over {\l^\a\l^\b}},\quad
(b_{\a\b\g}) = {{K^{[\a\b\g]}}\over {\l^\a\l^\b\l^\g}},\quad
(b_{\a\b\g\d}) = {{L^{[\a\b\g\d]}}\over {\l^\a\l^\b\l^\g\l^\d}}.}
It
is not difficult to show that \ghkl\ implies that $\{Q+\d, b\} = T$.
Using the methods of the previous subsection,
the corresponding globally defined Dolbeault form is therefore
\eqn\nonminb{b =   {{\lb_\a G^\a}\over{(\lb\l)}} +
{{\lb_\a r_\b H^{[\a\b]}}\over{(\lb\l)^2}}
 -
{{\lb_\a r_\b r_\g K^{[\a\b\g]}}\over{(\lb\l)^3}} -
{{\lb_\a r_\b r_\g r_\d L^{[\a\b\g\d]}}\over{(\lb\l)^4}}, }
which satisfies $\{Q_{nonmin}, b\} = T$. 

Finally, to construct
a $b_{nonmin}$ ghost satisfying $\{Q_{nonmin}, b_{nonmin}\} = T_{nonmin}$
where $T_{nonmin} = T + \bar w^\a \p\lb_\a - s^\a \p r_\a$, 
one defines $b_{nonmin} = b + s^\a \p\lb_\a$. Plugging in the explicit
form of the operators
$[G^\a, H^{[\a\b]},
K^{[\a\b\g]}, L^{[\a\b\g\d]}]$, one finds that \pst
\eqn\finalb{b_{nonmin} =
s^\a\p\lb_\a  + {{\lb_\a (2
\Pi^m (\g_m d)^\a-  N_{mn}(\g^{mn}\p\t)^\a
- J_\l \p\t^\a -{1\over 4} \p^2\t^\a)}\over{4(\lb\l)}}  }
$$+ {{(\lb\g^{mnp} r)(d\g_{mnp} d +24 N_{mn}\Pi_p)}\over{192(\lb\l)^2}}
-
{{(r\g_{mnp} r)(\lb\g^m d)N^{np}}\over{16(\lb\l)^3}} +
{{(r\g_{mnp} r)(\lb\g^{pqr} r) N^{mn} N_{qr}}\over{128(\lb\l)^4}}. $$

Note that throughout the rest of this paper, the subscript $nonmin$
will be dropped from $b_{nonmin}$ and $Q_{nonmin}$. Instead, 
we shall sometimes
use the subscript $min$, in order to stress the use of the minimal 
formalism. 

\subsec{Amplitude prescription}

Using the composite $b$ ghost defined in \finalb, the naive
topological prescription for $N$-point $g$-loop amplitudes is
\eqn\corrthree{A = \int_{{\CM}_{g, N}} d^{3g-3}\tau \Biggl\langle 
\prod_{j=1}^{3g-3}(\int dw_j \mu_j(w_j) b(w_j)) \prod_{r=1}^N
\ dz_r U(z_r) \Biggr\rangle}
where $\tau_j$ are the complex Teichmuller parameters, $\mu_j$ are
the associated Beltrami differentials, $\int dz U(z)$ are
the BRST-invariant integrated vertex operators which can
be assumed to be independent of the non-minimal fields, 
$\langle ~~\rangle$
denotes functional integration over the worldsheet fields, and
the right-moving contribution to $A$ is being ignored. 
Since
the ghost-number anomaly of the non-minimal formalism is
$+3$, this topological prescription is reasonable. 
However, as explained in \pst, there are two subtleties with this
amplitude
prescription which are associated with the functional integration
over the pure spinors.

The first subtlety is that the bosonic ghosts $(\l^\a,\lb_\a)$ have
22 non-compact zero modes, and integration over these
zero modes produces infinities when $\l\to\infty$. 
Similarly, the conjugate momenta
$(w_\a, \bar w^\a)$ have $22g$ non-compact zero modes on a genus
$g$ surface which also produce infinities when $w\to\infty$. Fortunately,
these infinities are cancelled by zeros coming
from integration over the zero modes of the fermionic variables
$(\t^\a, r_\a)$ and their conjugate momenta $(p_\a, s^\a)$.

The $0/0$ factors coming from
integration over the bosonic and fermionic zero
modes can be regularized by inserting an operator
${\cal N} = e^{ \{Q, \chi\}} $
into the integral over the zero modes. Since
${\cal N} = 1 + \{Q, \Omega\}$ for some $\Omega$, the choice
of $\chi$ does not affect BRST-invariant expressions.
A convenient choice for
$\chi$ is \pst
\eqn\conv{\chi =  -\lb_\a \t^\a - \sum_{I=1}^g (\half N_{mn}^I S^{mn I} 
+ J^I S^I  )}
where $[N_{mn}^I, J^I, S_{mn}^I, S^I]$ are the zero modes of
$[N_{mn}, J, S_{mn}, S]$ of \mingauge\ and \gaugeinvbar\
obtained by integrating these
currents around the $I^{th}$ $a$-cycle, e.g. $N_{mn}^I =
\oint_{a_I} dz N_{mn}(z)$. 

With this choice,
\eqn\calNloop{{\cal N} =
\exp \left( -\lb_\a\l^\a -r_\a \t^\a 
-\sum_{I=1}^g \Biggl[~\half N_{mn}^I \bar N^{mn I} + J^I \bar J^I 
+ {1\over 4} S_{mn}^I
~d^I\g^{mn}\l   + S^I
 ~\l^\a d^I_\a ~ 
\Biggr]~\right),}
which imposes an exponential cutoff for the non-compact bosonic
zero modes. Although ${\cal N}$ is not manifestly invariant under
spacetime supersymmetry transformations or under
modular transformations of the genus $g$ worldsheet,
it is easy to show that it changes by BRST-trivial quantities under
these transformations. Since ${\cal N}$ only involves worldsheet
zero modes, these BRST-trivial quantities are harmless and
cannot produce surface terms in the integral over the
Teichmuller moduli.
Note that the regulator ${\CN}$ is somewhat similar to the {\it
projection form} used in topological gauge theory to fix the
fermionic gauge invariance (see, e.g. \ref\cmr{S.~Cordes, G.W.~Moore, 
S.~Ramgoolam, {\it Lectures on 2-d Yang-Mills theory, 
equivariant cohomology and topological field theories}, hep-th/9411210}).

The second subtlety with \corrthree\ is more difficult
to resolve and comes from the singularities in the $b$ ghost of 
\finalb\ when $(\lb\l)\to 0$. Since the measure factor
for the pure spinors converges like $(\lb\l)^{11}$ when $(\lb\l)\to 0$,
these singularities are dangerous if they combine to diverge as fast
as $(\lb\l)^{-11}$. Since each $b$ ghost can diverge like
$(\lb\l)^{-3}$, there are potential problems with the amplitude
prescription when there are more than three $b$ ghosts, i.e. when
$g>2$. 

As explained in \pst, this second subtlety is related 
to the existence of the operator
\eqn\prob{\xi = {{\lb_\a\t^\a}\over{\l^\b \lb_\b + r_\b\t^\b}}
= (\lb\t) \sum_{n=1}^{11} {{(-r\t)^{n-1}}\over {(\l\lb)^n}} }
which satisfies $\{Q,\xi\}=1$ and diverges like
$(\l\lb)^{-11}$. Since $QV=0$ implies that 
$Q(\xi V)= V$, the existence of the operator $\xi$ naively implies that
the BRST cohomology is trivial. So if operators which diverge as
$(\l\lb)^{-11}$ are allowed in the Hilbert space, the BRST cohomology 
becomes trivial and one should expect to encounter problems in correlation
functions and scattering amplitudes.

It is instructive to give the {\cech} picture of the $\xi$-operator.
It is given by the inhomogeneous cochain:
\eqn\xico{{\bf \xi} = \left( {{\t}^{\a} \over {\l}^{\a} } \right) + 
\left( {{\t}^{\a}{\t}^{\b} \over {\l}^{\a}{\l}^{\b} } \right) + 
\ldots + \left( {{\t}^{1} \ldots {\t}^{16} \over {\l}^{1} \ldots {\l}^{16} } \right)}
which obeys:
\eqn\xicoc{{\d}{\bf \xi} + Q_{min} {\bf\xi}  = 1 \ .}
In the following section, this second subtlety will be resolved
by constructing a regularized version of the $b$ ghost which is
non-singular when $(\lb\l)\to 0$. After replacing the $b$ ghost
with its regularized version, it will be possible to
use the prescription of \corrthree\ to compute arbitrary multiloop 
amplitudes.

\newsec{Regularization of $b$ Ghost}

\subsec{Regularization of local operators}

In our regularization method,
we will deal with operators such as the $b$ ghost
which involve singular-looking expressions 
like 
$$
{1\over {( {\bar\l}{\l} )^{l}}}
$$
with $l<11$. It is important to show that correlation
functions of such operators are finite in the pure spinor $\b\g$ system. 
To this end we shall produce now a $Q$-invariant regularization, which
does not change the $Q$-cohomology class of an operator, while making
it explicitly non-singular. 

The idea of this regularization can be first explained in the example of
quantum mechanics, where we do not deal with the issue of $Q$-invariance.
So let us first study the quantum mechanics of a particle with zero
Hamiltonian in a phase
space with the coordinates $(p_m , q^m)$ for $m=1$ to $d$.
Suppose we face the following problem: 

Let ${\CO}_{l}(q)$ be a function which 
has a pole of order $l$ at the point $q=0$. 
Then, naively, the correlation function
\eqn\qmcorf{\langle {\CO}_{l_{1}} 
{\CO}_{l_{2}} \ldots {\CO}_{l_{p}} 
\rangle = \int {\dd}^{d}q \ {\CO}_{l_{1}}(q)  
{\CO}_{l_{2}}(q) \ldots {\CO}_{l_{p}} (q) }
is singular when $l_{1} + l_{2} + \ldots + l_{p} \geq d$.

Now imagine adding the Hamiltonian $\e^2 \Delta =
\e^2 g^{mn} p_{m}p_{n}$ where $\e$ is a constant. 
As long as the operators $\CO_{l_k}(q)$
are separated and satisfy the individual conditions
$ l_{k} < d $,
the smearing due to the heat kernel
evolution
will make them non-singular.
Indeed, we have
the heat kernel regularization of local operators: 
\eqn\hkrqm{{\CO}_{l}(q) \mapsto {\CO}_{l,\e} (q) = e^{\e^2\Delta} 
{\CO}_l (q)
= e^{- \e^2 g^{mn} p_m 
p_n} {\CO}_l(q) }
$$
= {1\over{(4\pi)^{d\over 2}}}
\int d^d f~ e^{- f^2} e^{i \e f^m p_m} {\CO}_l (q) 
= 
{1\over{(4\pi)^{d\over 2}}}
\int d^d f~ e^{- f^2} {\CO}_l (q +\e f)$$
$$= 
{1\over{(4\pi\e^2)^{d\over 2}}}
\int d^d q'~ e^{-{1\over\e}(q-q')^2} {\CO}_l (q')$$
where $q'=q+\e f$. Note that in the above derivation,
$g_{mn}$ is assumed to be constant
and the momenta $p_m$ are treated as operators.

Now, as long as $l  < d$, the integral in \hkrqm\ converges, and
is non-singular at $q = 0$:
\eqn\sngi{{\CO}_{l,\e}(q \to 0) \propto \e^{- l} < \infty.}
So the heat kernel regularization has ``smeared out''
the singularity of ${\CO}_{l} (q)$ at $q=0$.
A similar regularization will be now proposed for observables on 
the pure spinor space such that  
\eqn\hkr{{\CO}({\l}, {\bar\l}) \mapsto {\CO}_{\e} ({\l}, {\bar\l}) = 
e^{\e^2 {\Delta}} {\CO}({\l}, {\bar\l}).}

\subsec{ Regularization in pure spinor space}

Since $w_\a$ and $\bar w^\a$ are the conjugate momenta to $\l^\a$
and $\lb_\a$, a naive guess for the Laplacian on pure spinor space
is $\Delta = w_\a \bar w^\a$. So the naive generalization of 
\hkrqm\ to pure spinors is
\eqn\hkrpp{{\CO}_{\e} (\l,\lb) 
= {1\over{(4\pi)^{11}}}
\int d^{11} f d^{11} \bar f~ e^{-f^\a \bar f_\a} 
e^{i \e ( f^\a w_\a +\bar f_\a \bar w_\a)} 
{\CO} (\l,\lb) }
$$ 
= 
{1\over{(4\pi)^{11}}}
\int d^{11} f d^{11} \bar f~ e^{-f^\a \bar f_\a} 
{\CO} (\l',\lb') $$
where $\l'= \l+\e f$, 
$\lb'= \lb+\e \bar f$, and $f^\a$ and $\bar f_\a$ are pure spinors. 
However, since $\l' = \l+\e f$ is not necessarily
a pure spinor, this definition
needs to be modified.
The problem is that $w_\a$ and $\bar w^\a$ are not gauge-invariant 
under \gaugeone\ and \gaugebar, so their commutation relations 
with $\l^\a$ and $\lb_\a$ are not well-defined.

Gauge-invariant versions of $w_\a$ and $\bar w^\a$
can be defined as 
\eqn\defW{W_\a = (\l\bar f)^{-1} (-{1\over 8} 
(\l\g^{mn} w)(\g_{mn}\bar f)_\a
- {1\over 4}(\l w) \bar f_\a ),}
$$\bar W^\a = (f\lb)^{-1} (-{1\over 8} (\lb\g^{mn} \bar w)(\g_{mn}f)^\a
-{1\over 4} (\lb \bar w) f^\a )$$
where $f^\a$ and $\bar f_\a$ are constant pure spinors. Using the identity
\eqn\ident{\d_\b^\g \d_\a^\d =
{1\over 2} \g^m_{\a\b}\g_m^{\g\d} - {1\over 8} (\g^{mn})_\a{}^\g 
(\g_{mn})_\b{}^\d  -{1\over 4} \d_\a^\g \d_\b^\d}
which can be proven by contracting both sides of \ident\ with 
$\g_p^{\a\b}$, $\g_{pqr}^{\a\b}$ or $\g_{pqrst}^{\a\b}$, 
one finds that 
\eqn\identwo{
(\l \bar f) w_\a = {1\over 2}(w\g^m\bar f)(\g_m \l)_\a -{1\over 8}
(\l\g^{mn} w)(\g_{mn}\bar f)_\a -{1\over 4} (w\l)\bar f_\a.}
So in the gauge $w\g^m \bar f= \bar w \g^m f=0$, $W_\a=w_\a$
and $\bar W^\a = \bar w^\a$.

Note that $W_{\a}$ can be written more compactly as:
\eqn\wal{W_{\a} = {1\over 4} ({N\kern -.1in \slash}{\l}^{-1}_{\a} - 
J {\l}^{-1}_{\a})}
where
\eqn\invspin{{\l}^{-1} = {1\over {\l}{\bar f}} {\bar f}}
Although the global gauge-invariant differential operators on $\CP$ are polynomials in $N_{mn}$ and $J$, which act by ``rotations" which 
preserve the point ${\l} = 0$, the parameters of the ``rotations" in 
\wal\invspin\ are singular at ${\l}=0$, allowing the 
operators $W_{\a}$ to shift the ``bad" point ${\l}=0$ in what follows. 

One can therefore define
\eqn\hkrrr{{\CO}_{\e} (\l,\lb) =
 {1\over{(4\pi)^{11}}}
\int d^{11} f d^{11} \bar f~ e^{-f^\a \bar f_\a} 
e^{i \e ( f^\a W_\a +\bar f_\a \bar W^\a)} 
{\CO} (\l,\lb) }
as a gauge-invariant version of \hkrpp. Although $W_{\a}$ of
\defW\ needs to be normal-ordered, the normal-ordering ambiguity commutes
with $\l^\a$ and therefore does not affect the definition of \hkrrr.
Using the OPE's of $N^{mn}$ and $J$ with $\l^\a$, $\CO_\e(\l\lb)$
can be expressed as: 
\eqn\hkrss{{\CO}_{\e} (\l,\lb) =
{1\over{(4\pi)^{11}}}
\int d^{11} f d^{11} \bar f~ e^{-f^\a \bar f_\a} 
{\CO} (\l',\lb')}
where 
\eqn\twoprime{ {\l'}^\a = 
e^{i \e f^\b W_\b } \l^\a 
= \l^\a + \e f^\a - 
{{[(\l + \e f)\g^m (\l+ \e f)] (\g_m \bar f)^\a} 
\over{4(\l+\e f)^\b \bar f_\b}}, }
$${\lb'}_\a = 
e^{i \e  \bar f_\b \bar W^\b } \lb^\a =  
\lb_\a
 + \e \bar f_\a - 
{{[(\lb + \e\bar f)\g^m (\lb+\e\bar f)] (\g_m f)_\a} 
\over{4(\lb+\e \bar f)_\b {f}^\b}}.$$ 
Note that $\int d^{11}f d^{11} \bar f$ denotes $\int \Omega\bar\Omega
(f^\a \bar f_\a)^3$ where $\Omega= {{ d^{11}f}\over {f^3}}$ and 
$\bar\Omega= {{ d^{11}\bar f}\over {\bar f^3}}$ are the holomorphic
and antiholomorphic top-forms on 
the space of pure spinors. As will be seen in the following subsection,
the additional factor of $(f^\a \bar f_\a)^3$
in the integration measure is absent in the BRST-invariant
generalization of \hkrss.

It is easy to check that $\l'$ and $\lb'$ of
\twoprime\ are pure spinors satisfying 
$\l'\g_m\l'= \lb'\g_m \lb'=0$. In fact, one way to derive \twoprime\
is to require that $\l'$ is a pure spinor and
that ${\l'}^\a = \l^\a + \e f^\a + \Omega^m (\g_m \bar f)^\a$
for some $\Omega^m$. The additional term proportional to $\Omega^m$ comes
from the commutation relation 
\eqn\commu{[W_\a, \l^\b] = \d_\a^\b -\half
 (\g^m \l)_\a (\g_m{\l}^{-1})^\b.}
Another way to understand \twoprime\ is to use the parameterization 
\eqn\usepa{
{\l}^{\a} = {\l}^{+} \left( 1 , u_{ab} ,  u_{[ab}u_{cd]}
 \right)}
of a pure spinor. Then, 
given two pure spinors, ${\l}$ and ${\e}f ={\e}f^{+} 
\left( 1 , {\phi}_{ab}, {\phi}_{[ab}{\phi}_{cd]} \right)$, one 
can construct the third one
by taking
\eqn\twoprimes{{\l}^{\prime\a}  = \left( {\l}^{+} + {\e}f^{+} \right)
 \cdot \left( 1, u'_{ab} , u'_{[ab}u'_{cd]} \right)}
where
\eqn\addspin{u'_{ab} = {{\l}^{+} u_{ab} + {\e}f^{+} {\phi}_{ab}    
\over {\l}^{+} + {\e}f^{+}}.} 
This ``addition of pure
spinors" is equivalent to \twoprime\  for ${\bar f} = ( 1, 0, 0)$. 

It will now be argued that 
${\cal O}_\e(\l,\lb)$ in \hkrss\ is well-defined for all values of 
$\l$ if one assumes that 
${\cal O}(\l',\lb') \sim (\l'\lb')^{-n}$
where $0\leq n< 11$.
Firstly, note that
when $\l^\a \to 0$ and $\lb_\a\to 0$, \twoprime\ implies that
${\l'}^\a \to f^\a$ and ${\lb'}_\a \to \bar f_\a$. So as in 
the quantum-mechanical example, the regularization
${\CO}(\l,\lb) \mapsto {\CO}_{\e} (\l,\lb) $
smears out the singularity of ${\CO}(\l,\lb)$ at $\l=\lb=0$.
Since
${\CO}(\l,\lb)$ diverges slower than $(\l\lb)^{-11}$, there
are no singularites in 
${\CO}_\e (\l,\lb)$ when $\l=\lb=0$.

Secondly, note that when $(\l+\e f)^\b \bar f_\b \to 0$, \twoprime\
implies that $\l'$ diverges. 
However, since 
${\cal O}(\l',\lb') \sim (\l'\lb')^{-n}$ for $n\geq 0$, 
${\CO}_\e (\l,\lb)$  remains finite when $\l'$ diverges.

Finally, suppose that $\l$ is chosen such that $\l^{\prime\a}=
e^{i\e f^\b W_\b}\l^\a$ vanishes. For \hkrss\ to be well-defined, it
is necessary that the measure factor $d^{11} f d^{11}\bar f$
converges faster than $(\l'\lb')^n$ when $\l'\to 0$. It will be useful to
consider separately the cases
when $\l^\a=0$ and when $\l^\a$ is non-zero.
When $\l^\a=0$, \twoprime\ implies that $\l'= \e f$.
So $d^{11}f d^{11}\bar f =\e^{22} d^{11} \l' d^{11}\lb'$, which
converges as $(\l'\lb')^{11}$ near $\l'=0$.
When $\l^\a$ is non-zero,
one can choose a Lorentz
frame in which $\l^+$ is non-zero and $u_{ab}=0$ in \usepa.
Using the parameterization of \twoprimes, $\l^{\prime\a}\to 0$ implies that
$(\l^+ + \e f^+) \to 0$ with $u'_{ab}$ held fixed. Since \addspin\
implies that $\phi_{ab} \to (\l^+ + \e f^+)  {{u'_{ab}}\over{\e f^+}}$,
one
finds that $d^{11} f d^{11}\bar f$ converges like $|\l^+ + \e f^+|^{22}$ when
$\l^{\prime\a}\to 0$, which is fast enough to cancel the $(\l'\lb')^{-n}$
divergence if $n<11$.

\subsec{BRST-invariant regulator}

To make the regularization method of \hkrrr\ BRST-invariant, 
it is convenient to introduce constant bosonic pure spinors
$f^\a$ and $\bar f_\a$, and constant constrained fermions $g^\a$
and $\bar g_\a$, satisfying
\eqn\constra{  g\g^m f =0,\quad \bar g \g^m \bar f=0,
\quad f\g^m f =0, \quad  \bar f\g^m\bar f=0,}
and to define $[f^\a, \bar f_\a, g^\a, \bar g_\a]$ to transform
under BRST transformations as 
\eqn\brstc{ [Q, f^\a] =0, \quad
[Q, \bar f_\a] = \bar g_\a, \quad
\{Q, g^\a\} = f^\a, \quad
\{Q, \bar g_\a\} = 0.}
The constraints of \constra\ imply that 
$f^\a$, $\bar f_\a$, $g^\a$ and $\bar g_\a$ each have eleven
independent components.

Geometrically, the $Q$-operator \brstc\ can be identified with the operator
$$
{\pb} + \iota_{E}
$$
acting on the space ${\Omega}^{\bullet, \bullet} ({\CP})$
of all differential forms on the space of pure spinors.
Here $E = {\l} {{\p}\over {{\p}{\l}}}$ is the holomorphic Euler vector field. The familiar
$U(1)$ action on $\CP$ is generated by the vector field
$U = - i ( E - {\bar E})$. In this picture $g^{\a} = df^{\a}$, ${\bar g}_{\a} = d{\bar f}_{\a}$. 
The  operator \brstc\ can be viewed as
a ``half'' of the $U(1)$ equivariant 
differential ${\dd} + \iota_{U} \sim Q + {\bar Q}$.

{}We have
$$
Q ( {\bar f}_{\a} g^{\a} ) = {\bar f}_{\a} f^{\a} + {\bar g}_{\a} g^{\a}
$$
which is the $U(1)$-equivariant symplectic form on $\CP$.  
Furthermore, if one defines 
\eqn\defwa{W_\a = 
(\l\bar f)^{-1} \left(  {1\over 4} N_{mn} (\g^{mn} \bar f)_\a
 -{1\over 4} J \bar f_\a \right)\ , }
$$
V^\a = 
(f\lb)^{-1} ( {1\over 4} S_{mn} (\g^{mn} f)^\a
 - {1\over 4} S f^\a ),$$
one finds that 
\eqn\defqwa{[Q, W_\a] = 
(\l\bar f)^{-1} \left( {1\over 8} (\l \g_{mn}d) (\g^{mn} \bar f)_{\a}
 +{1\over 4} (\l d) \bar f_\a +
{1\over 4} N_{mn} (\g^{mn} \bar g)_\a
 -{1\over 4} 
J \bar g_\a \right) - {{(\l\bar g)}\over{ (\l \bar f)}} W_\a, }
$$\{ Q, V^{\a} \} = 
(f\lb)^{-1} \left( {1\over 4} \bar N^{mn} (\g^{mn} f)^\a
 -{1\over 4} \bar J f^\a \right) - {{(f r)}\over{(f\lb)}} V^{\a} \ . $$

Up to terms involving fermions, it is easy to verify that
$f^\a W_\a + \bar f_\a \bar W_\a = Q( g^\a W_\a + \bar f_\a V^\a)$.
Therefore, a BRST-invariant generalization of \hkrrr\ is
\eqn\hkrqq{ 
{\CO}_\e (\l,\lb)
= 
\int {\dd}^{11} f {\dd}^{11} \bar f  {\dd}^{11} g {\dd}^{11} \bar g
~ e^{- (\bar f_\a f^\a + \bar g_\a g^\a)} 
e^{i \e Q( g^\a W_\a + \bar f_\a V^\a)} 
{\CO} (\l,\lb) \ ,}
or, in a more concise way:
\eqn\hkrqqq{{\CO}_{\e} ( {\l}, {\lb} ) = 
\int_{\CP} e^{- {\bar f}_{\a}f^{\a} + d{\bar f}_{\a} \wedge d f^{\a}}
e^{i {\e} Q( df^\a W_\a + \bar f_\a V^\a)} 
{\CO} (\l,\lb) \ .}
The integration measure in \hkrqqq\ is defined by simply
expanding the exponential until the top degree form, i.e. the $22$-form,
is produced. So in the BRST-invariant version of $\CO_\e(\l,\lb)$, 
the integration measure is simply $\Omega\bar\Omega\Sigma\bar\Sigma$
where 
\eqn\topf{\Omega = {{d^{11}f}\over{f^3}},\quad
\bar\Omega = {{d^{11}\bar f}\over{\bar f^3}},\quad
\Sigma = f^3 d^{11}g,\quad
\bar\Sigma = \bar f^3 d^{11}\bar g,}
are the top degree forms.

As before, one
can show 
that ${\CO}_\e(\l,\lb)$ is well-defined at $\l=\lb=0$ as long
as 
${\CO} (\l,\lb) $ diverges slower than $(\l\lb)^{-11}$.
And since ${\CO}_\e = {\CO} + \{ Q, \chi_\e\}$ for some $\chi_\e$,
BRST-invariant amplitudes involving ${\CO}_\e$ will be independent
of the parameter $\e$.

\subsec{Regularized $b$ ghost}

The regularization method of \hkrqq\
is easily generalized to the worldsheet
operator $b(z)$ of \finalb\ by defining
\eqn\regulb{
b_\e(y) = 
\int {\dd}^{11} f {\dd}^{11} \bar f {\dd}^{11} g {\dd}^{11} \bar g
~ e^{-(\bar f_\a f^\a + \bar g_\a g^\a)} ~b'(y)}
where
\eqn\defbprime{b'(y) = 
e^{i \e \{Q,~ g\oint dz U(z) +  \bar f \oint dz V(z)\}} 
~b(y) ~
e^{- i \e \{Q,~ g\oint dz U(z) +  \bar f \oint dz V(z)\}},  }
$U_\a(z)$ and $V^\a(z)$ are the holomorphic currents defined in
\defwa, and the contour integrals in \defbprime\ go around the point $y$.

Since $b_\e(y) = b(y) + \{Q, \chi_\e(y)\}$ for some $\chi_\e(y)$,
$\{Q, b_\e(y)\} = T(y)$ and BRST-invariant
scattering amplitudes are independent
of the value of $\e$. Furthermore, since $b(y)$ diverges slower than
$(\l\lb)^{-11}$, $b_\e(y)$ has no singularities at $\l(y)=\lb(y)=0$.

\newsec{Multiloop Amplitude Prescription}

Substituting the regularized $b_\e$ ghost of \regulb\ for
the $b$ ghost, 
the $N$-point $g$-loop amplitude prescription of \pst\ becomes 
\eqn\regamp{A = \lim_{\e\to 0} \int d^{3g-3 }\tau
\ \Biggl\langle 
\prod_{j=1}^{3g-3} 
(\int dw_j \mu_j(w_j) b_\e (w_j)) 
\prod_{r=1}^N \int dz_r U_r(z_r) 
~{\cal N} \Biggr\rangle }
where ${\cal N}$ is the same regulator for the zero modes as
defined in \calNloop.
For non-zero $\e$, the functional integral is well-defined and,
since 
$b_\e = b + \{Q, \chi_\e\}$ for some $\chi_\e$, the amplitude
prescription is independent of $\e$ up to possible surface terms. So one
is free to take the limit $\e\to 0$ after performing the functional
integral.

In the multiloop amplitude prescription of \regamp, the functional integral
is vanishing unless the integrand contributes 16 $\t$ zero modes for open
superstrings, or 32 $\t$ zero modes for closed Type II superstrings.
Since the $b_\e$ ghost is manifestly
spacetime supersymmetric, these $\t$ zero modes can only come either from 
superfields in the
external vertex operators $U_r$ or from the 
$e^{-(\l\lb+ r \t)}$ term
in the zero mode regulator ${\cal N}$ of \calNloop.

To evaluate \regamp, it is useful to separate the correlation
function into two types of terms: terms in which at least one $\t$ zero
mode comes from the zero mode regulator ${\cal N}$, and terms in which
none of the $\t$ zero modes come from the zero mode regulator.
As will now be explained, the first type of terms can contribute to
F-terms in the ten-dimensional effective action and are easier to
evaluate since they do not require $\e$-regularization. The second type 
of terms are more complicated to evaluate,
however, it will be shown that they only contribute near the region
$\l=\lb=0$.

\subsec{Ten-dimensional F-terms}

Although one does not know how to construct off-shell D=10 superspace
actions, 
one can construct higher-derivative
D=10 superspace actions which are functions of on-shell
linearized superfields.
Ten-dimensional F-terms are defined as manifestly gauge-invariant
terms in the superspace effective action which cannot be written
as integrals over the maximum number of $\t$'s. 
In the massless 
vertex operator for open superstrings,
the gauge-invariant superfield of lowest dimension is $W^\a(x,\t)$
whose lowest component is the gluino of dimension $1\over 2$. Since
N=1 D=10 superspace contains 16 $\t$'s, any term in the superspace
action involving $M$ superfields $W^\a$ which is integrated over
the full superspace has dimension $\geq (M+16)/2.$ Therefore, any
term in the N=1 D=10 superspace action involving $M$ field-strengths
which has
dimension less than $(M+16)/2$ is necessarily an N=1 D=10 F-term.

In the massless vertex operator for closed Type II superstrings, 
the gauge-invariant
superfield of lowest dimension is $W^{\a\b}(x,\t,\tb)$ whose lowest
component is the Ramond-Ramond field strength of dimension 1. Note that
the dilaton and axion are dimension zero fields, but they always appear
with derivatives in the massless vertex operator. Since N=2 D=10
superspace contains 32 $\t$'s, 
any term in the superspace
action involving $M$ superfields $W^{\a\b}$ which is integrated over
the full superspace has dimension $\geq (M+16).$ Therefore, any
term in the N=2 D=10 superspace action involving $M$ field-strengths 
which has
dimension less than $(M+16)$ is necessarily an N=2 D=10 F-term.
For example, since the curvature tensor $R_{mnpq}$ has dimension 2,
the term 
\eqn\arg{\int d^{10}x \sqrt{g} \p^L R^M}
in the Type II effective
action is an N=2 D=10 F-term if $L+2M< M+16$, i.e. if $L+M<16$.

If all $\t$ zero modes come from superfields in
the external vertex operators in \regamp,
the resulting term in the superspace effective action is expressed as an
integral over the maximum number of $\t$'s and therefore does not contribute
to F-terms. However, if 
any of the $\t$ zero modes come from ${\cal N}$,
the resulting term in the superspace effective action is expressed
as an integral over a subset of the $\t$'s. Although this does not
automatically imply that it is an F-term (since it may be possible
to rewrite the expression as an integral over all the $\t$'s), it might
contribute to F-terms. 

So any term in the scattering amplitude 
which contributes to an F-term in the effective action 
must receive at least one $\t$ zero mode from ${\cal N}$.
It will now be shown that any such term diverges slower than
$(\l\lb)^{-11}$ and therefore does not require $\e$-regularization
of the $b$ ghost.

To show that terms receiving $\t$ zero modes from ${\cal N}$ do
not require $\e$-regularization, first note that BRST invariance
implies that the 
$e^{-(\l\lb+ r \t)}$ term in ${\cal N}$ can be modified to
$e^{-\rho(\l\lb+ r \t)}$ for any positive $\rho$. Because
\eqn\argrho{e^{-\rho(\l\lb+ r \t)} = 
e^{\{Q, -\rho \t\lb\}} = 1 + \{Q,\xi_\rho\} }
for some $\xi_\rho$, BRST-invariant ampitudes are independent
of the value of $\rho$. 

Suppose one computes the amplitude $\langle F(\l,\lb) ~{\cal N}\rangle$
where
$F(\l,\lb)$ is some BRST-invariant operator. Then $\rho$-independence
implies that the
$(-\rho \t r)^n$ terms in 
$$e^{-\rho(\l\lb+ r \t)} 
= e^{-\rho\l\lb} \left( 1 + \sum_{n=1}^{11} {1\over n!
} (-\rho \t r)^n \right)$$ 
can only contribute to $\langle F(\l,\lb)~ {\cal N}\rangle$
if $\int d^{11}\l d^{11}\lb~ F(\l,\lb) ~e^{-\rho\l\lb}$ has poles in $\rho$.
But this implies that $F(\l,\lb)$ diverges slower than $(\l\lb)^{-11}$
since 
\eqn\divergs{\int d^{11}\l d^{11}\lb~ (\l\lb)^{-l} ~ e^{-\rho\l\lb}
\propto \rho^{l-11}.}

So $\t$ zero modes in ${\cal N}$ can only contribute to
$\langle F(\l,\lb)~ {\cal N}\rangle$
if $F(\l,\lb)$ diverges slower than $(\l\lb)^{-11}$, which implies
that $\e$-regularization is unnecessary. So any term which receives
$\t$ zero modes from ${\cal N}$ can be evaluated by directly setting
$\e=0$ before performing the correlation function.

\subsec{Terms requiring $\e$-regularization}

For terms in which
all $\t$ zero modes come from the vertex operators, $\rho$-independence
of the amplitude implies that \divergs\ cannot have poles in $\rho$,
so $l\geq 11$. Therefore, 
$\int d^{11}\l d^{11}\lb~ (\l\lb)^{-l}$ diverges and $\e$-regularization
of the $b$ ghost is necessary. Although the computation of
these terms is
complicated, integration
over the non-minimal fermions
$r_\a$ will imply that the only contribution to these terms
comes from the region near
$\l=\lb=0$.

To show that the only contribution come from the region near
$\l=\lb=0$, first note that the unregularized $b$ ghost
of \finalb\ commutes with the conserved charges
\eqn\charges{\oint dz (r_\a s^\a - \l^\a w_\a) \quad{\rm and}\quad
\oint dz \lb_\a s^\a.}
In other words, 
all terms in the unregularized $b$ ghost have $r$-charge opposite to their
$\l$-charge, and are invariant under the shift
$\d r_\a = c\lb_\a$ for constant $c$.
Furthermore, since \divergs\ has no poles in $\rho$, $\rho$-independence
implies that one can directly
set $\rho=0$ in ${\cal N}$ so that
\eqn\calNzero{{\cal N}_{\rho=0} =
\exp\left(~
\sum_{I=1}^g \Biggl[~-\half N_{mn}^I \bar N^{mn I} - J^I \bar J^I ~~
- {1\over 4} S_{mn}^I
~d^I\g^{mn}\l  ~~ - S^I
 ~\l^\a d^I_\a ~ 
\Biggr]~\right).}
One can check that \calNzero\ also commutes with \charges, so if
the vertex operators $U_r$ are chosen to be independent of the non-minimal
variables, the unregularized integrand
\eqn\unregi{
\prod_{s=1}^{3g-3} b(w_s) 
\prod_{r=1}^N U_r (z_r) 
~{\cal N}_{\rho=0}}
commutes with the charges of \charges.

This implies that before performing $\e$-regularization of these terms, 
the integrand of \regamp\ has the form
\eqn\formi{\sum_{k\geq 0}
C_k^{\a_1 ... \a_{11+k}} ~~{{r_{\a_1} ... r_{\a_{11+k}}}
\over {(\lb \l)^{11+k}}} }
where $C_k^{\a_1 ... \a_{11+k}}$ are operators which carry
zero $\l$-charge and zero $r$-charge, and which satisfy
$\lb_{\a_1} C^{(\a_1 ... \a_{11+k})}_k =0$.

Since $r_\a$ has eleven zero modes, at least $k$ of 
the $(11+k)$ $r$'s in \formi\ must contribute non-zero modes. Furthermore,
when $k=0$, at least one of the eleven $r$'s in \formi\
must contribute a non-zero
mode because of the invariance under $\d r_\a= c\lb_\a$.
So for the correlation function to be non-vanishing, terms coming from
the $\e$-regularization must provide non-zero modes of $s^\a$ which can
contract with these non-zero modes of $r_\a$.

These $s_\a$ non-zero modes  
can come from $V^\a$ of 
\defwa\ through the regularization factor
$e^{i\e \bar g_\a \oint dz V^\a}$ in $b_\e$, 
which means that each $s^\a$ non-zero mode
comes multiplied
by a factor of $\e$. So these terms vanish in the limit $\e\to 0$,
except near $\l=\lb=0$ where $\e$-regularization can produce poles in $\e$.
Therefore, to evaluate terms in which all $\t$ zero modes come from
vertex operators, one only needs to evaluate the functional integral
$\int d^{11}\l d^{11}\lb$ near the point $\l=\lb=0$. It might be
possible to explicitly evaluate the contributions of these delta
functions at $\l=\lb=0$, however, this will not be attempted here.

\vskip 15pt
{\bf Acknowledgements:} We would like to thank M.~Green, H.~Verlinde and 
E.~Witten for discussions, and O.~Bedoya Delgado and C.~R.~Mafra
for correcting a coefficient in equation (3.7).  

The research of NB was partially supported by 
CNPq grant 300256/94-9, Pronex
grant 66.2002/1998-9, and FAPESP grant 04/11426-0, and that of NN by
European RTN under the contract 005104  "ForcesUniverse", by ANR 
under the grants ANR-06-BLAN-3$\_$137168 and ANR-05-BLAN-0029-01, and by the grants {\cyr RFFI} 06-02-17382
and {\cyr NSh}--8065.2006.2. We both 
thank the Institute for Advanced Study at Princeton for hospitality during the
work on this project. NB also thanks 
IHES, Bures-sur-Yvette, for hospitality and NN thanks MSRI, 
Department of Mathematics at UC Berkeley and
NHETC at Rutgers University for hospitality
during various stages of preparation of the manuscript.

\footatend\vfill\supereject\immediate\closeout\rfile\writestoppt
\baselineskip=14pt\centerline{{\bf References}}\bigskip{\frenchspacing%
\parindent=20pt\escapechar=` \input refs.tmp\vfill\eject}\nonfrenchspacing

\bye